\documentclass[acmsmall,screen]{acmart}

\AtBeginDocument{%
  \providecommand\BibTeX{{%
    \normalfont B\kern-0.5em{\scshape i\kern-0.25em b}\kern-0.8em\TeX}}}

\usepackage{syntax}
\usepackage{graphicx}
\usepackage{booktabs}
\usepackage{subcaption}
\usepackage{mathtools}
\usepackage{dsfont}
\usepackage{paralist}
\usepackage{caption}
\usepackage{amscd}
\usepackage{tikz}
\usetikzlibrary{matrix}
\usetikzlibrary{fit}
\usetikzlibrary{automata}
\usetikzlibrary{arrows,shapes,backgrounds,positioning,calc,decorations.pathmorphing}
\usepackage{multirow}
\usepackage{dashbox}
\usepackage{amsfonts}
\usepackage[normalem]{ulem}
\usepackage{wrapfig}
\usepackage{appendix}
\usepackage{algpseudocode}
\usepackage[linesnumbered,ruled,noend]{algorithm2e}
\usepackage{amsmath} 
\algnewcommand\algorithmicforeach{\textbf{for each}}
\algdef{S}[FOR]{ForEach}[1]{\algorithmicforeach\ #1\ \algorithmicdo}
\usepackage{listings}
\usepackage[utf8]{inputenc}

\usepackage[T1]{fontenc}
\usepackage{textcase}
\usepackage{empheq}
\usepackage{tcolorbox}

\usepackage{xspace}

\newcommand{\Nats}{\ensuremath{\mathbb{N}}\xspace}
\newcommand{\trans}[3]{\ensuremath{#1 \xrightarrow{#2} #3}\xspace}

\newcommand{\conf}{\mathbf{q}}
\newcommand{\altconf}{\mathbf{p}}
\newcommand{\wqo}{\preceq}
\newcommand{\altwqo}{\trianglelefteq}

\newcommand{\eg}{e.g.\xspace}
\newcommand{\ie}{i.e.\xspace}

\newcommand{\LS}{S}
\newcommand{\ls}{s}

\newcommand{\guard}{G}
\newcommand{\guardset}{\mathcal{\guard}}
\newcommand{\free}{free\xspace}
\newcommand{\freech}{independent\xspace}

\newcommand{\cutoffgsp}{\text{\emph{Cutoff}}_{Core}\xspace}
\newcommand{\cutoffch}{\text{\emph{Cutoff}}_{\merc}\xspace}

\newcommand{\senderset}{\hat{\ls}\xspace}

\newcommand{\lemref}[1]{Lemma~\ref{lem:#1}}

\newcommand{\secref}[1]{Sec.~\ref{sec:#1}}

\newcommand{\figref}[1]{Fig.~\ref{fig:#1}}

\newcommand{\algoref}[1]{Algo.~\ref{algo:#1}}
\newcommand{\thmref}[1]{Theorem~\ref{thm:#1}}

\newcommand{\appref}[1]{App.~\hyperref[app:#1]{\ref*{app:#1}}}

\newcommand{\tabref}[1]{Table~\ref{table:#1}}
\newcommand{\lineref}[1]{Line~\ref{line:#1}}

\newcommand{\para}[1]{\noindent{\bf \em #1}}

\newcommand{\mtt}[1]{\mathtt{#1}}
\newcommand{\Nat}{\mathbb{N}}

\newcommand\tuple[1]{\langle #1 \rangle}

\newcommand{\assign}{$\coloneq$}
\newcommand{\true}{{\tt true}}

\newcommand{\n}{n}

\newcommand{\Q}{locs}
\newcommand{\LQ}{S}
\newcommand{\GQ}{Q}

\newcommand{\q}{\locname}

\newcommand{\loq}{s}
\newcommand{\gq}{q}

\newcommand{\consagree}{agreement\xspace}

\newcommand{\GActions}{\mathcal{A}}

\newcommand{\V}{vars}
\newcommand{\acts}{acts}
\newcommand{\T}{T}

\newcommand{\GT}{R}

\newcommand{\actname}{\text{\termb{act}}}
\newcommand{\consnamep}{\text{\termb{part}}}
\newcommand{\consnamev}{\text{\termb{cons}}}
\newcommand{\locname}{\text{\termb{loc}}}

\newcommand{\M}{\mathcal{M}}

\newcommand{\PG}{PG}  

\newcommand{\indicatorr}[1]{\mathds{1}(#1)} 

\newcommand{\ch}{\textsc{agree}\xspace}

\newcommand{\sysname}{QuickSilver}
\newcommand{\kinarach}{\textsc{\sysname}\xspace}
\newcommand{\kinarachtitle}{{\sc \sysname}\xspace}

\newcommand{\gqe}{\gq_{\mtt{end}}}

\newcommand{\chset}{\text{\termb{pcpt}}}	
\newcommand{\chlocsv}{\Q^\consnamev}	
\newcommand{\chlocsp}{\Q^\consnamep}	
\newcommand{\chcard}{k}			
\newcommand{\chvar}{\text{\termb{pVar}}}	

\newcommand{\Ind}{I_n}  
\newcommand{\gchset}{S}	
\newcommand{\gqs}{\gq_{\mtt{start}}}
\newcommand{\winset}{W}	

\newcommand{\aqs}{ \overline{\mathcal{\M}}}	

\newcommand{\supp}{\mathsf{supp}\xspace}

\newcommand{\gbc}{GSP\xspace} 

\newcommand{\hlm}{\gbc model\xspace}	
\newcommand{\ml}{\M_{\merc}}	
\newcommand{\mhp}{\M_{Core}}
\newcommand{\absfn}{\alpha}	

\newcommand{\pl}{P_{\merc}} 
\newcommand{\ph}{P_{\gbc}} 
\newcommand{\php}{P_{\textsc{Core}}}
\newcommand{\Mh}{\M_{\gbc}} 
\newcommand{\permissible}{permissible\xspace} 
\newcommand{\Permissible}{Permissible\xspace} 
\newcommand{\amenable}{cutoff-amenable\xspace}

\newcommand{\wellbehaved}{well-behaved\xspace}

\newcommand{\pairwise}{rendezvous\xspace} 
\newcommand{\Pairwise}{Rendezvous\xspace} 

\newcommand{\pld}{payld}
\newcommand{\ad}{payldDom\xspace}

\newcommand{\dslvariables}{variables\xspace}

\newcommand{\dslactions}{actions\xspace}

\newcommand{\dslsend}{sendrz}
\newcommand{\dslatmost}{atmost}
\newcommand{\dslbroadcast}{sendbr}
\newcommand{\dslcommStmt}{sendStmt\xspace}
\newcommand{\machine}{process\xspace}

\newcommand{\ValueCons}{\terma{Consensus}\xspace}
\newcommand{\PartitionCons}{\terma{Partition}\xspace}

\newcommand{\RewriteValueConsTransitions}{RewriteConsensusTransitions}

\newcommand{\PartitionConsDSL}[3]{\colorit{\terma{Partition}\textbf{<}#1\textbf{>(}#2\textbf{,}#3\textbf{)}}}

\newcommand{\entry}{$\_$} 

\newcommand{\colorit}[1]{\colorbox{cyan!15}{#1}}

\definecolor{Gray}{gray}{0.85}
\definecolor{LightRed}{rgb}{1,0.7,0.7}
\definecolor{LightGreen}{rgb}{0.7,1,0.7}

\newcommand{\wam}{partition-and-move\xspace}

\newcommand{\valstore}{value-consensus\xspace}

\definecolor{realGreen}{rgb}{0.0, 0.5, 0.0}

  \newcommand{\neww}[1]{\textcolor{black}{#1}}
    \newcommand{\newww}[1]{\textcolor{black}{#1}}

\newcommand{\mnx}[1]{}

\newcommand{\aml}{\textsc{Mercury}\xspace}
\newcommand{\merc}{\textsc{Merc}\xspace}
\newcommand{\amlexpansion}{Modeling Event Reaction and Coordination Using symmetRY\xspace}

\newcommand{\phase}{phase\xspace}

\newcommand{\phases}{phases\xspace}

\newcommand{\chwellbehaved}{phase-compatible\xspace} 
\newcommand{\chwellbehavedness}{phase-compatibility\xspace} 
\newcommand{\Chwellbehavedness}{Phase-compatibility\xspace} 
\newcommand{\CHwellbehavedness}{Phase-Compatibility\xspace}

\newcommand{\amenability}{cutoff-amenability\xspace} 

\newcommand{\AMenability}{Cutoff-Amenability\xspace}

\newenvironment{claim}[1]{\par\noindent{\emph{Claim:}}\space#1}{}
\newenvironment{claimproof}[1]{\par\noindent{\emph{Proof:}}\space#1}{\hfill \qed}

\newcommand{\fch}{\mathcal{F}_{\merc}} 
\newcommand{\fgsp}{\mathcal{F}_{\textsc{Core}}}

\newcounter{sarrow}

\newcommand{\aprimitive}{agreement primitive\xspace}
\newcommand{\aprimitives}{agreement primitives\xspace}
\newcommand{\Aprimitives}{Agreement Primitives\xspace}

\newcommand{\sigmaOf}[1]{\sigma(#1)}
\newcommand{\valOf}[2]{#1.\sigmaOf{#2}}

\renewcommand{\syntleft}{$\langle$\normalfont\itshape} 
\renewcommand{\syntright}{$\rangle$}

\newcommand{\optintvar}{optIntVar}

\newcommand{\nontDSL}[1]{\syntleft #1\syntright}
\newcommand{\nont}[1]{$\langle$\textit{#1}$\rangle$}
\newcommand{\termb}[1]{\code{#1}}
\newcommand{\terma}[1]{\textbf{\code{#1}}}

\newcommand{\Pid}{PID\xspace}
\newcommand{\pid}{PID\xspace}
\newcommand{\pids}{PIDs\xspace}
\newcommand{\decVal}{decVar}

\newcommand{\sendrzid}[4]{#1 $ $\xrightarrow{\text{\terma{sendrz}}(#2,#3)}$ $#4}
\newcommand{\sendrz}[3]{#1$ $\xrightarrow{\text{\terma{sendrz}}(#2)} $ $ #3}
\newcommand{\sendbr}[3]{#1$ $\xrightarrow{\text{\terma{sendbr}}(#2)} $ $ #3}
\newcommand{\recvrz}[3] {#1$ $\xrightarrow{\text{\terma{recvrz}}(#2)} $ $ #3}
\newcommand{\recvbr}[3] {#1$ $\xrightarrow{\text{\terma{recvbr}}(#2)} $ $ #3}

\newcommand{\vc}{VC_{\consnamev}} 

\newcommand{\pcw}{\code{win:} PC_{\consnamep}}
\newcommand{\pcl}{\code{lose:} PC_{\consnamep}}

\newcommand{\ourskip}{\smallskip}

\newcommand{\act}{\text{\termb{e}}}
\newcommand{\actb}{\text{\termb{f}}}

\newcommand{\gsactions}{E_\mathtt{{global}}}
\newcommand{\pwactions}{E_\mathtt{{rend}}}
\newcommand{\relatedIntPw}{\mathcal{R}}
\newcommand{\relatedIntPwph}{\relatedIntPw_{\pha}}
\newcommand{\initPhases}{\text{\emph{inPhases}}}
\newcommand{\expandedPhases}{\text{\emph{exPhases}}}
\newcommand{\pha}{ph}

\newcommand{\states}{st}

\newcommand{\spec}[3]{\phi_{#1,#2}(#3)}
\newcommand{\mpvp}{MPVP\xspace}

\newcommand{\gspp}{\aml \textsc{Core}\xspace}
\newcommand{\crash}{\mathtt{crash}}

\newcommand{\crashstate}{\loq_{cr}}
\newcommand{\failset}{F}

\newcommand{\firable}{initiable\xspace} 

\newcommand{\newwqo}{\precsim}
\newcommand{\failstop}{crash-stop\xspace}

\newcommand{\noncrashed}{non-crashed\xspace}

\definecolor{bluekeywords}{rgb}{0.13,0.13,1}
\definecolor{greencomments}{rgb}{0,0.5,0}
\definecolor{turqusnumbers}{rgb}{0.17,0.57,0.69}
\definecolor{redstrings}{rgb}{0.5,0,0}

\definecolor{qualifiers}{rgb}{0.5,0,0.63}
\definecolor{types}{rgb}{0.5,0,0}
\definecolor{communication}{rgb}{0.5,0,0}
\definecolor{categories}{rgb}{0.5,0,0.63}
\definecolor{controlstmts}{rgb}{0.5,0,0}
\definecolor{handlermeta}{rgb}{0.13,0.13,1}
\definecolor{setops}{rgb}{0.5,0,0}
\definecolor{cons}{rgb}{0.13,0.13,1}

\definecolor{hlcolor}{rgb}{0.13,0.13,1}

\lstdefinelanguage{FSharp}
    {keywords=[1]{env, initial}, 
    keywordstyle=[1]\bfseries\color{qualifiers},
    keywords=[2]{int, idset, id, bool, unit},
    keywordstyle=[2]\bfseries\color{types}, 
    keywords=[3]{sendrz, sendbr,recv,br,rz}, 
    keywordstyle=[3]\bfseries\color{communication},
    keywords=[4]{machine, variables, actions, location, invariant,process},
    keywordstyle=[4]\bfseries\color{categories},
    keywords=[5]{if,else,goto},
    keywordstyle=[5]\bfseries\color{controlstmts},
    keywords=[6]{on, win, lose, passive, do, where, atmost},
    keywordstyle=[6]\bfseries\color{handlermeta},
    keywords=[7]{add, remove, contains,decVar,winS,loseS,payld},
    keywordstyle=[7]\bfseries\color{setops},
    keywords=[8]{PartitionCons, ValueCons},
    keywordstyle=[8]\bfseries\color{cons},
    sensitive=false,
    morecomment=[l][\color{greencomments}]{///},
    morecomment=[l][\color{greencomments}]{//},
    morestring=[b]",
    stringstyle=\color{redstrings}
    }

    \lstnewenvironment{dsl}
  {
    \lstset{
        language=FSharp,
        basicstyle=\ttfamily,
         numbers=left,
    	stepnumber=1,
        breaklines=true,
        escapeinside={{(*}{*)}},
        columns=fullflexible
        }
        }
  {
  }

    \lstnewenvironment{udsl}
  {
    \lstset{
        language=FSharp,
        basicstyle=\ttfamily,
        breaklines=true,
        escapeinside={{(*}{*)}},
        columns=fullflexible
        }
        }
  {
  }

    \lstnewenvironment{sdsl}
  {
    \lstset{
        language=FSharp,
        basicstyle=\ttfamily\footnotesize,
         numbers=left,
    	stepnumber=1,
        breaklines=true,
        escapeinside={{(*}{*)}},
        columns=fullflexible
        }
        }
  {
  }

      \lstnewenvironment{ssdsl}
  {
    \lstset{
        language=FSharp,
	 basicstyle=\ttfamily\scriptsize,
         numbers=left,
    	stepnumber=1,
        breaklines=true,
        escapeinside={{(*}{*)}},
        columns=fullflexible
        }
        }
  {
  }

     \lstnewenvironment{sudsl}
  {
    \lstset{
        language=FSharp,
        basicstyle=\ttfamily\tiny,
        breaklines=true,
        escapeinside={{(*}{*)}},
        columns=fullflexible
        }
        }
  {
  }

   \newenvironment{newtext}{\color{black}}{\ignorespacesafterend}

   \newenvironment{newtextt}{\color{black}}{\ignorespacesafterend}   

\newcommand{\code}[1]{\texttt{#1}} 
\newcommand{\reducespace}{-1pt}
\newcommand{\squeezecaption}{\vspace{\reducespace}}
\newcommand{\reallyreducespace}{-10pt}
\newcommand{\reallysqueezecaption}{\vspace{\reallyreducespace}}
\setcopyright{rightsretained}
\acmJournal{PACMPL}
\acmYear{2021} \acmVolume{5} \acmNumber{OOPSLA} \acmArticle{157} \acmMonth{10} \acmPrice{}\acmDOI{10.1145/3485534}

\begin{document}

\title{\kinarachtitle: Modeling and Parameterized Verification for Distributed Agreement-Based Systems} 



\author{Nouraldin Jaber}
\affiliation{%
  \institution{Purdue University}
  \city{West Lafayette}
  \country{USA}}
\email{njaber@purdue.edu}

\author{Christopher Wagner}
\affiliation{%
	\institution{Purdue University}
	\city{West Lafayette}
	\country{USA}}
\email{wagne279@purdue.edu}

\author{Swen Jacobs}
\affiliation{%
	\institution{CISPA Helmholtz Center for Information Security}
	\city{Saarbr\"{u}cken}
	\country{Germany}}
\email{jacobs@cispa.saarland}

\author{Milind Kulkarni}
\affiliation{%
	\institution{Purdue University}
	\city{West Lafayette}
	\country{USA}}
\email{milind@purdue.edu}

\author{Roopsha Samanta}
\affiliation{%
	\institution{Purdue University}
	\city{West Lafayette}
	\country{USA}}
\email{roopsha@purdue.edu}


\begin{abstract}

	The last decade has sparked several valiant efforts in deductive verification of distributed agreement protocols such as consensus and leader election. 
	Oddly, there have been far fewer verification efforts that go beyond the core protocols and target applications that are {\em built on top of} agreement protocols. 
	This is unfortunate, as agreement-based distributed services such as data stores, locks, and ledgers are ubiquitous and potentially permit modular, scalable verification approaches that mimic their modular design. 
	
	We address this need for verification of distributed agreement-based systems through our novel modeling and verification framework,
	\kinarach, that is not only modular, but also fully automated. The key enabling feature of \kinarach is our encoding of {\em abstractions of verified agreement protocols} that facilitates modular, decidable, and scalable automated verification. 
	We demonstrate the potential of \kinarach by modeling and efficiently
	verifying a series of tricky case studies, adapted from real-world
	applications, such as a data store, a lock service, a  surveillance system,  a  pathfinding algorithm for mobile robots, and more.


\end{abstract}

\ccsdesc[500]{Theory of computation~Program verification}
\ccsdesc[500]{Theory of computation~Distributed computing models}
\ccsdesc[500]{Theory of computation~Automated reasoning}
\ccsdesc[500]{Theory of computation~Verification by model checking}
\ccsdesc[300]{Theory of computation~Abstraction}
\ccsdesc[300]{Theory of computation~Concurrency}
\ccsdesc[300]{Theory of computation~Program analysis}

%
\keywords{Parameterized Verification, Modular Verification, Distributed Systems}

\maketitle
\section{Introduction}
\label{sec:intro}

Modern distributed services such as data stores, logs, caches, queues, locks, and ledgers heavily rely on {\em distributed agreement} to perform their higher-level functions---processes in these distributed services need to agree on a leader, on the members of a group, on configurations, or on owners of locks. Notable instances of such {\em distributed agreement-based} services include the Chubby lock service~\cite{burrows2006chubby} and RedisRaft key-value store~\cite{redisraft}, which are built on top of the Paxos~\cite{lamport1998part} and Raft~\cite{ongaro2014search} consensus algorithms, respectively. 
The importance of agreement protocols as a key building block in distributed services has
sparked significant verification efforts for these protocols~\cite{Padon.PaxosMadeEPR.OOPSLA.2017,chand2016formal,lamport2002specifying,cousineau2012tla,Druagoi.PSyncPartiallySynchronous.POPL.2016,Maric.CutoffBoundsConsensus.X.2017,Liu.ClarityEfficiencyDistributed.X.2012,Woos.PlanningChangeFormal.X.2016,Druagoi.LogicbasedFrameworkVerifying.X.2014,Perez.PaxosConsensusDeconstructedAbstracted.X.2018}.
%
%
%
%
\neww{Intriguingly, with rare exceptions, these efforts restrict their attention to the core protocols and do not consider the distributed services that {\em build} on those protocols. This is unfortunate because there are arguably more distributed systems that build on agreement protocols than there are implementations of these core protocols. Moreover, such agreement-based systems are more likely to be developed by non-experts who can benefit from verification.} 
%
%
%
In this paper, we ask {\em can we develop modular modeling and verification frameworks for distributed agreement-based systems} by (1) assuming that the underlying agreement protocols are verified separately and (2) encapsulating their complexities within cleanly-defined abstractions? 
Such an approach would both allow us to leverage the heroic efforts towards verifying agreement protocols as well as ease the burden of modeling the distributed systems that rely on those protocols.
\neww{We note that} existing verification efforts for agreement-based systems that go beyond core protocols~\cite{Hawblitzel.Ironfleet.SOSP.2015,Gleissenthall.Pretend.Synchrony.POPL.2019,Liu.ClarityEfficiencyDistributed.X.2012,Padon.IvySafetyVerification.PLDI.2016}, with the exception of~\cite{Sergey.ProgrammingProvingDistributed.POPL.2017,TLC.Griffin.ICFP.2020}, do not leverage the availability of verified agreement artifacts through systematic agreement abstractions.

We further ask: can our agreement abstractions enable {\em fully automated, parameterized verification} for interesting classes of agreement-based systems? This second question is an open one. The 
parameterized model checking problem (PMCP)---the problem of algorithmically verifying correctness 
of systems parameterized by the number of processes---is well-known to be undecidable in its full generality~\cite{kozan.pmcp.undecidable.1986,Suzuki.PMCP.UndecidableFirstPPR.1988.PPR}.
While decidability has been shown for some restricted classes of distributed systems, it is unclear whether agreement-based systems allow for a decidable parameterized verification procedure at all.
Past verification efforts for agreement protocols/implementations as well as agreement-based systems sidestep the decidability issue by  
preferring  the use of interactive or semi-automated deductive verification over model checking. 
The appeal of {\em push-button} verification that does not require a user to provide inductive invariants or manipulate a theorem prover, however, remains undeniable. 
We argue that abstracting away and separately verifying the intricate details of agreement (using deductive techniques) should yield {\em simpler} models of agreement-based systems that may {\em now} become amenable to decidable and scalable model checking. 

In this paper, we propose the \kinarach framework for modeling 
and parameterized model checking of distributed agreement-based systems. 
\kinarach advances a brand new verification strategy for agreement-based systems that is not only modular, but also fully automated.

\subsection*{The \kinarach Framework}

In our design of \kinarach, we address several questions: 

\begin{compactenum}
\item \textit{How should we abstract agreement?}
The primitives we develop to abstract agreement must be sufficiently general to capture the essential characteristics of a wide variety of agreement protocols, while still permitting decidable parameterized model checking of distributed systems with such primitives.

\item \textit{How should we model our systems?}
The modeling language we use for distributed agreement-based systems 
should match the manner in which system designers build their programs.

\item \textit{How should we identify systems that enable decidable and scalable verification?} 
The fundamental obstacle we need to tackle is the undecidability of PMCP.
Thus, we must find {\em easily-checkable} conditions under which the verification of systems we model (including their use of agreement primitives) is decidable. Further, because our goal is to model fairly complex systems, we must endeavor to find scalable approaches for verifying these systems.
\end{compactenum}
In particular, \kinarach makes the following contributions.
\smallskip

\para{\aml: A Modeling Language with Agreement Primitives.}
We carefully examined a range of agreement protocols in the literature, such as consensus
and leader election, and observed that while the protocol internals differed substantially, their externally-observable behavior could be captured with two {\em agreement primitives} \neww{(namely, \PartitionCons and \ValueCons)} that have simple semantics and abstract away the protocols' implementation details. \neww{The \PartitionCons primitive allows a set of participant processes to divide themselves into groups (e.g., leaders and followers). The \ValueCons primitive allows its participants, with each proposing a value, to agree on a finite set of decided values.} \secref{prelim} presents a new, intuitive modeling language, \aml\footnote{\amlexpansion}, that allows designers to model \neww{finite-state} distributed systems using these agreement primitives, and hence design systems without worrying about the internals of the core agreement protocols.
\smallskip

\label{propertiesIntro}
\para{Parameterized Verification of \aml Systems.} With \aml's primitives abstracting away the messy details of distributed agreement, we observe that the resulting higher-level systems can be more amenable to automated verification.
\secref{verification} identifies a broad class of \aml systems that permit decidable and efficient parameterized verification. In particular, we present two key results. First, we identify syntactic conditions on \aml systems that yield decidability of PMCP. Second, we identify additional syntactic conditions that, for a given class of safety\neww{/reachability} properties, enable {\em practical} parameterized verification by providing {\em cutoffs}: a number $k$ of processes such that verifying the correctness of a {\em fixed-size} $k$-process system implies the correctness of arbitrary-sized systems. This result means that {\em non-parameterized} model checkers can be leveraged to provide parameterized verification.

\neww{We prove both results by (1) defining \gspp, a novel extension of the decidable and cutoff-yielding fragments of a recently proposed abstract model for distributed systems~\cite{GSP} and (2) showing that \aml systems satisfying our syntactic conditions are {\em simulation equivalent} to systems in \gspp.}  

\neww{The class of safety/reachability properties to which these results apply include properties forbidding the reachability of global states where {\em more than a fixed number} of processes are simultaneously in some pivotal local states. An example of such a property is mutual exclusion of a certain critical local state, i.e., {\em no more than two} processes can reach the critical state simultaneously in any execution. These results currently do not apply to liveness properties, e.g., a leader is eventually elected, or to arbitrary safety properties, \eg, there exists {\em at least one leader} at all times, or, {\em no more than half of the processes} can simultaneously be leaders.}\ourskip 


\para{Implementation, \aml Benchmarks, and Evaluation.} 
With our decidability and cutoff results for \aml programs in hand, we have an approach that enables scalable, parameterized verification of distributed agreement-based systems 
{\em in theory.}
\secref{evaluation} instantiates the theoretical results of \kinarach by presenting an implementation of a cutoff-driven parameterized verification procedure for \aml systems.
Crucially, \kinarach \emph{automatically} checks the syntactic conditions that yield practical parameterized verification. 
When a system is not practically verifiable, \kinarach provides best-effort feedback suggesting modifications to the system that may make it so.
We show that complex distributed agreement-based systems including a data store, a lock service, a  surveillance system,  a  pathfinding algorithm for mobile robots, the Small Aircraft Transportation System (SATS) protocol~\cite{satsref}, and several other interesting applications can be naturally and succinctly modeled in \aml, and can then be efficiently verified.

\subsection{Related Work}
\label{sec:related}

We first compare with the most related lines of work.  
\ourskip

\para{Global Synchronization Protocols (GSPs).} 
In recent work, \citet{GSP} propose  a new model, GSP, for crash- and failure-free distributed systems and present decidability and cutoff results for parameterized verification of systems in the model. This model supports global transitions associated with global guards which can be used by multiple processes to synchronize collectively and simultaneously. Such global transitions and guards can be used, {\em in theory}, to carefully encode abstractions of agreement protocols. However, the GSP model is  an abstract, theoretical model based on counter abstraction and does not provide an  intuitive, accessible interface for system designers; for instance, processes in the GSP model cannot use local variables and are specified as low-level state-transition systems with {\em manually-inferred} guards. Additionally, users are required to {\em manually check} if their GSP system models fall within the decidable, cutoff-yielding fragment.

In contrast, \kinarach (i) provides a user-friendly modeling language for distributed systems with inbuilt  primitives that are designed to abstract agreement protocols, (ii) supports process \failstop failures, (iii) pushes the boundaries of decidable parameterized verification by expanding the GSP decidability fragment, and (iv) includes a {\em fully-automated implementation} for checking if \aml programs belong to the expanded decidable, cutoff-yielding fragment.

\para{Modular Verification with Abstract Modules.} Disel \cite{Sergey.ProgrammingProvingDistributed.POPL.2017} and TLC \cite{TLC.Griffin.ICFP.2020} leverage the same
observation we do---that distributed applications build on standard protocols---and enable users to incorporate abstractions of such protocols to provide modular verification using the Coq theorem prover.
The user is responsible for providing both the
high-level descriptions of the underlying protocols as well as the inductive invariants
needed to link protocols to their clients and/or enable
{\em horizontal composition} with other protocols. 
The TLC framework could potentially reason about agreement-based systems as it supports {\em vertical composition}, but the user would need to manually incorporate abstractions of the underlying agreement protocols. 
In contrast, \kinarach is equipped with intuitive, inbuilt primitives that abstract \consagree protocols and facilitate vertical composition and fully-automated parameterized verification.
\ourskip


In what follows,  we discuss other broad themes of verification approaches for distributed systems.\ourskip

\para{Semi-Automated, Deductive Verification.} 
Approaches for semi-automated, deductive verification of distributed 
protocols and implementations expect a user to specify inductive invariants~\cite{Aneris.2020,Rahli.InterfacingProofAssistants.X.2012,Wilox.Verdi.PLDI.2015,Sergey.ProgrammingProvingDistributed.POPL.2017,Wilcox.ProgrammingLanguageAbstractions.X.2017,Doenges.VerificationImplementationsDistributed.X,Woos.PlanningChangeFormal.X.2016,Padon.IvySafetyVerification.PLDI.2016,andersen2019distributed,Feldman.PhaseStructures.CAV.2019}. 
Some approaches~\cite{Padon.PaxosMadeEPR.OOPSLA.2017,Padon.ReducingLivenessSafety.POPL.2017,Taube.ModularityDecidabilityDeductive.PLDI.2018,damian.communicationclosed.CAV.2019}
enable more (but not full) automation by translating the user-provided system
and inductive invariants into a decidable fragment of first-order logic (e.g.,
effectively propositional logic (EPR)~\cite{EPR}) or a model with a
semi-automatic verification procedure (e.g., the Heard-Of model~\cite{HOmodel}). 
Recent work~\cite{Gleissenthall.Pretend.Synchrony.POPL.2019,kragl2020inductive} proposes the use of Lipton's reduction~\cite{LiptonReduction} to reduce reasoning about asynchronous programs to synchronous and sequential programs, 
respectively, thereby greatly simplifying the invariants needed. 
Our approach builds on deductive verification for agreement protocols to 
enable modeling and automated parameterized verification of systems built on 
top of verified agreement protocols.\ourskip 

\para{Model Checking.} 
Prior work on PMCP identifies decidable fragments based on restrictions on the communication primitives, specifications, and structure of the system \cite{emerson2003model,GS92,AminofKRSV18,AusserlechnerJK16,DelzannoRB02,GSP,EsparzaFM99}. 
To enable efficient parameterized verification, prior work additionally identifies cutoff results for various classes of systems, e.g., cache coherence protocols \cite{EK03a}, guarded protocols \cite{Jacobs.AnalyzingGuardedProtocols.X.2018}, consensus protocols \cite{Maric.CutoffBoundsConsensus.X.2017}, and self-stabilizing systems \cite{Bloem.SynthesisSelfstabilisingByzantineresilient.X.2016}. Unfortunately, no existing decidability and cutoff results, except for those in~\cite{GSP}, extend to agreement-based systems.

There has also been some work on model checking and synthesis of distributed systems with a
\emph{fixed} number of finite-state
processes~\cite{Alur.AutomaticSynthesisDistributed.X.2017,Alur.AutomaticCompletionDistributed.X.2015,Alur.SynthesizingFinitestateProtocols.X.2014,Liu.ClarityEfficiencyDistributed.X.2012,Yang.MODISTTransparentModel.X.2009,Damm.AutomaticCompositionalSynthesis.X.2014}. However, these frameworks  are not naturally extendable for parameterized reasoning and do not 
consider abstractions of agreement protocols for improving scalability of verification in the fixed-size setting.\ourskip

\section{\kinarach Overview}
\label{sec:overview}

This section presents an illustrative example of a complex system that leverages multiple instances of distributed agreement for its high-level function.
It then provides an overview of the key building blocks of our
modeling language and verification approach using the example.
We begin the section with a brief review of distributed agreement protocols.
\ourskip

\para{Distributed Agreement Protocols.}
Distributed  agreement protocols enable a set of distributed participants, each proposing one value, to {\em collectively decide} on a set of proposals in the presence of failures and asynchrony. There are many variants of agreement protocols with small differences in their decision objectives. For instance, the participants may wish to decide on a single proposal~\cite{lamport1998part,Mencius,FastPaxos},
an infinite sequence of proposals~\cite{ongaro2014search,chandra2007paxos}, or a finite set of leaders amongst themselves~\cite{bullyalgo,improvedBullyAlgo}. Despite these variations, any \emph{correct} agreement protocol is characterized by the following three guarantees~\cite{Lynch:1996:DA:525656}: (i) \emph{agreement}---all participants decide on the same set of proposals, (ii) \emph{validity}---every proposal in the decided set of proposals must have been proposed by a participant, and (iii) \emph{termination}---all participants eventually decide. Accordingly, recent work in verification of agreement protocols  and/or their implementations focuses on guaranteeing agreement, validity, and termination (or, some reasonable variant of these properties).  

\subsection{Illustrative Example: Distributed Store} 
\label{sec:illustrative}

\begin{figure*}[ht]
\centering
\begin{tcolorbox}[colback=white,sharp corners,boxrule=0.3mm,top=1mm,bottom=1mm]
\hspace{1em}\begin{minipage}{0.47\textwidth}
\begin{ssdsl}
process DistributedStore
variables 
    int[1,5] cmd (*$\coloneq$*) 1
    int[1,2] stored (*$\coloneq$*) 1	
actions 
  env 
    rz doCmd : int[1,5]
    rz ackCmd : int[1,5]
    rz ret : int[1,2]
    br LeaderDown : unit 	

initial location Candidate (*\label{line:candidate}*)
  on (*\PartitionConsDSL{elect}{All}{1}*)  (*\label{line:elect}*)
    win: goto Leader
    lose: goto Replica
		
location Leader (*\label{line:leader}*)
  on recv(doCmd) do (*\label{line:docmd}*)
    cmd (*$\coloneq$*) doCmd.payld (*\label{line:docmd-start}*)
    if(cmd <= 2 && stored != cmd)
      goto RepCmd
    else if(cmd = 3) (*\label{line:read}*)
      sendrz(ret[stored], doCmd.sID)
    else 
      goto RepCmd  (*\label{line:docmd-end}*)
\end{ssdsl}
\end{minipage}
\begin{minipage}{0.70\textwidth}
 \lstset{firstnumber=26}
\begin{ssdsl}
location RepCmd
  on (*\colorit{\terma{\ValueCons}\textbf{<}vc\textbf{>}\textbf{(}All\textbf{,}1\textbf{,}cmd\textbf{)}}*) do(*\label{line:vc1}*)
    cmd (*$\coloneq$*) vcCmd.decVar[1]
    if(cmd <= 2) /*set*/
      stored (*$\coloneq$*) cmd
    else if(cmd = 4) /*inc*/
      stored (*$\coloneq$*) stored + 1
    else /*dec*/
      stored (*$\coloneq$*) stored - 1
    sendrz(ackCmd[cmd], doCmd.sID)
    goto Leader

location Replica (*\label{line:replica}*)
  on (*\colorit{\terma{\ValueCons}\textbf{<}vc\textbf{>}\textbf{(}All\textbf{,}1\textbf{,}_\textbf{)}}*) do(*\label{line:vc2}*)
    cmd (*$\coloneq$*) vcCmd.decVar[1]
    if(cmd <= 2) /*set*/
      stored (*$\coloneq$*) cmd
    else if(cmd = 4) /*inc*/
      stored (*$\coloneq$*) stored + 1
    else /*dec*/
      stored (*$\coloneq$*) stored - 1
  on recv(LeaderDown) do  (*\label{line:leaderdown}*)
    goto Candidate
\end{ssdsl}
\end{minipage}
\end{tcolorbox}

\begin{tcolorbox}[colback=white,sharp corners,boxrule=0.3mm, top=1mm, bottom = 2.5mm]
\neww{\footnotesize{
    \begin{compactenum}[\textbf{Safety Property:}]
    \item In every reachable state, there is at most one leader. 
    \item In every reachable state, all processes in locations \termb{Replica} and \termb{Leader}
          agree on the value of the variable \termb{stored}.
    \end{compactenum}
}}
\end{tcolorbox}

\caption{\aml Representation of a Distributed Store Process Definition. 
	A process in a \aml program consists of a collection of variables, communication
	actions, and locations with associated event handlers. Each event handler
	consists of an event and a reaction to that event. An event is an empty event,
	a receive of a communication action, or one of the two \aprimitives:
	\terma{\PartitionCons} and \terma{\ValueCons}. Reactions typically consist of
	a block of update statements, control statements, and/or sends of
	communication actions.}
\label{fig:casestudycode} 
\end{figure*}

%

Suppose a system designer wants to model and verify a distributed store where multiple processes consistently replicate and update a piece of stored data in response to client requests. The clients may request various operations on the stored data including read and update. To ensure that the data is consistent across all replicas, the designer decides to use distributed agreement protocols to determine which operation these replicas should execute next. For efficiency, they use a leader election protocol to pick a leader that acts as a single point of contact to handle requests from the clients. These requests are replicated to all other processes using a consensus algorithm to maintain consistent stored data throughout the system. This design pattern is common in distributed services like fault-tolerant key-value stores (e.g.,~\citet{redisraft}).
The safety properties for this system are: (1) there is at most one leader at any given time and (2) all processes agree on the stored data. 

Notice that the designer's scheme for the distributed store uses different agreement protocols as building blocks and is inherently {\em modular}. Moreover, the safety properties are about the high-level design of the distributed store and do not refer to the internals of the leader election and consensus protocols used. Hence, it is sensible to also adopt a modular approach to reasoning about the correctness of the design. Specifically, 
instead of reasoning about the distributed store as a monolithic program with all agreement protocols modeled explicitly, one can assume that the underlying agreement protocols are verified separately and verify the system where these protocols are replaced with simpler abstractions that capture their behaviors. 

Our framework, \kinarach, and its modeling language \aml (presented in \secref{prelim})
 enables the designer to utilize such a modular verification approach with the agreement protocols represented using special primitives (denoted \PartitionCons and \ValueCons for leader election and consensus, respectively) that soundly abstract the semantics of agreement protocols. 

For instance, the designer may model their distributed store in \aml as shown in  (\figref{casestudycode}). 
Processes start in the \code{Candidate} location (\lineref{candidate}) and coordinate with each other  (\lineref{elect}) to elect one leader to move to the \code{Leader} location (\lineref{leader}), while the remaining processes become replicas and go to the \code{Replica} location (\lineref{replica}). The leader can receive requests from clients (via the \code{doCmd} message on \lineref{docmd}), while the replicas wait for the leader to replicate requests to them.

\neww{When the leader receives a request from a client (which could be one of several potential commands), it handles the request on lines~\ref{line:docmd-start}--\ref{line:docmd-end}.} A command payload consists of either a directive to set the value to $1$ or $2$ (\code{cmd <= 2}); to read the stored value (\code{cmd = 3}); or to increment (\code{cmd = 4}) or decrement (\code{cmd = 5}) the stored value.
 On a read request, the leader responds by returning its stored data to that client (\lineref{read}). When the leader receives any update request, such as a request to set, increment, or decrement its stored data, the leader moves to the \code{RepCmd} location to initiate a round of consensus to replicate the operation to all replicas in the system (\lineref{vc1} for the leader, \lineref{vc2} for the replicas). When consensus is complete, all processes update their stored data by executing the operation on which they have agreed, and the leader returns an acknowledgment message to the requesting client. 
 \neww{In the event that the leader process has crashed (modeled by the environment sending  a special \code{LeaderDown} message, not shown in \figref{casestudycode}), all processes in the \code{Replica} location receive that message (\lineref{leaderdown}) and return to the \code{Candidate} location so a new leader may be elected.} 
 

 

 

	Note that the key feature of \aml's design arises from the encapsulation of the two distributed agreement operations that occur: choosing a leader (\lineref{elect}), captured by the \PartitionCons primitive, and that leader's replicating commands to the replicas (lines~\ref{line:vc1} and~\ref{line:vc2}), captured by the \ValueCons primitive. These primitives have carefully-designed semantics that capture the essence of agreement protocols, and allow \aml programs to be built without considering how that agreement is implemented. \secref{primitives} describes these primitives and motivates their design in more detail.

Having represented the distributed store in \aml, the designer can now utilize \kinarach's push-button parameterized verifier to check the correctness of any distributed system consisting of one or more such identical processes with respect to the two safety properties. 
In particular, \kinarach can automatically verify that this \aml distributed store  satisfies the safety properties regardless of the number of processes, in less than a minute. 
Moreover, any refinement of this program that instantiates the \PartitionCons and \ValueCons primitives with some verified leader election or consensus protocol, respectively, is also guaranteed to satisfy the safety properties.

\subsection{Agreement Primitives}
\label{sec:primitives}

The design of \aml's agreement primitives is driven by our goal of
automated, parameterized verification of
agreement-based systems. Thus, the granularity of abstraction in the agreement
primitives was carefully chosen to strike a balance between (a) capturing the {\em
	essence} of most practical agreement protocols without modeling
protocol-specific behavior and (b) facilitating 
decidability of PMCP (when combined with additional syntactic conditions). To meet these objectives, we propose {\em two}
agreement primitives, \PartitionCons and \ValueCons, that can individually model two 
common variants of agreement that we refer to as \emph{\wam} \consagree and \neww{\emph{\valstore}} \consagree, respectively. The two primitives can further be composed together to model other variants of agreement. We informally explain these primitives here, and provide a more formal treatment of their semantics in \secref{choosemodel}. 
\ourskip


\para{The \PartitionCons Primitive.} The \PartitionCons agreement primitive is used to
model \wam \consagree, where a set of participants wishes to partition
itself into groups.  Instances of \wam \consagree include variants of leader election
protocols 
 which partition the participants into two groups: leaders (or, {\em winners})
and non-leaders (or, {\em losers}). Note that each participant essentially proposes 
their {\em process index (\pid)}, which, in a parameterized 
distributed system with an unbounded number of processes, is drawn from an infinite domain.
To enable decidability of PMCP for systems that use
\wam \consagree protocols, the cardinality of exactly one group must be unbounded (e.g., non-leaders), while that of all other groups must be finite (e.g., leaders). This disallows, for instance, partitioning the participants into two equal sets of winners and losers. Fortunately, we observe that most \wam agreement protocols pick a finite number of winners that is independent of the number of participants. 

Each \PartitionCons \consagree primitive has an identifier, and takes two parameters: the set of participants and the desired number of winners. 
The reaction of each \PartitionCons primitive contains a \terma{win} (resp. \terma{lose}) handler that indicates how the process behaves upon winning (resp. losing).

\neww{\noindent{\em Example.} Distributed Store (\figref{casestudycode}) uses \wam agreement in \lineref{elect}, modeled using a handler in the \code{Candidate} location with
the agreement event: \PartitionCons\termb{<elect>(All,1)}. 
This event uses the \PartitionCons primitive with identifier \termb{elect} and is used to pick 1 process out of the set of all processes (\termb{All}). The winners (resp. losers) of this agreement instance move to location \code{Leader} (resp. \code{Replica}).} \ourskip

\para{The \ValueCons Primitive.} The \ValueCons agreement primitive is used to model
\valstore \consagree, where a set of participants, each proposing one value,
wishes to decide on (a set of) values. Instances of \valstore \consagree
include protocols such as Paxos~\cite{lamport1998part}, Fast
Paxos~\cite{FastPaxos}, and 
Mencius~\cite{Mencius}.
To enable decidability of PMCP for systems that use
\valstore \consagree, we restrict the domain of the
proposed values to be finite. We note that many distributed systems aim to solve coordination-like problems rather than compute a function over their data. Hence, infinite concrete data domains can soundly be treated as finite abstract data domains using, \neww{for example}, predicate abstraction. 

Each \ValueCons primitive has an identifier, and takes three parameters: the set of participants, the number of proposals to be decided and (an optional) variable that a process uses to propose a value. 




\neww{\noindent{\em Example.} Distributed Store (\figref{casestudycode}) uses \valstore agreement in Lines \ref{line:vc1} and \ref{line:vc2} to allow the processes to decide which operation should be executed next. In particular, the processes use two handlers: one in the \code{RepCmd} location with the agreement event \ValueCons\termb{<vc>(All,1,cmd)}, and another in the \code{Replica} location with the agreement event \ValueCons\termb{<vc>(All,1,_)}. Both primitives have the identifier \code{vc} and allow all processes to participate (\code{All}). In the former, the leader proposes a value from the variable \termb{cmd}, while in the latter, the replicas propose no value (denoted by _). The primitive decides on one value that can be accessed by all participants using the expression \termb{vc}\terma{.\decVal}\termb{[}1\termb{]}.}

\smallskip

\para{Composition of Primitives.} The \PartitionCons and \ValueCons
primitives can be composed to model \consagree protocols like
Multi-Paxos~\cite{chandra2007paxos} and Raft~\cite{ongaro2014search}, where a set of
participants wish to decide on a potentially infinite sequence of values. Instead of invoking \consagree on every value of the sequence individually, such protocols enhance practicality by first electing a leader that proposes the values, while the rest of the processes accept such values.
Such protocols can be modeled by using a \PartitionCons primitive to elect a
leader, and then using \ValueCons primitives to have the leader propose values
in subsequent rounds. Our Distributed Store example uses such a composition: upon receiving a \termb{doCmd} request, the leader uses the \ValueCons primitive to agree with the replicas. Note that the replicas pass an empty proposal (denoted _) as only the leader should be proposing values.
\smallskip




\subsection{Parameterized Verification in \kinarach}
\label{sec:pcmp-overview}
Since PMCP is a well-known undecidable
problem, it is not immediately obvious if parameterized verification is even decidable
for \aml programs with agreement primitives. To this
end, we first identify an expanded decidable fragment, \gspp, of an existing abstract model of distributed systems. Furthermore, we present two additional theoretical results that enable decidable and efficient parameterized verification for \aml programs. These results are based on \neww{establishing} a correspondence between \aml programs and \neww{programs in the more abstract \gspp model}, and appealing to \gspp's decidability and cutoff results.


\smallskip

\para{Decidable Parameterized Verification.} We identify syntactic
conditions, called {\em \chwellbehavedness conditions}, on systems with
agreement primitives that yield decidability of PMCP. 
Informally, the \chwellbehavedness conditions capture systems that proceed in
{\em phases}: each process is always in the same phase as every other process
and all processes, simultaneously, move from one phase to the next using some
global synchronization such as synchronous broadcasts or
agreement. The \chwellbehavedness conditions ensure that the system's ability
to move between phases is independent of the number of processes, thereby 
paving the way towards decidable parameterized verification. 

\noindent{\em Example.} Distributed Store (\figref{casestudycode}) is \chwellbehaved and hence enables decidable parameterized verification. The system starts in a phase where all processes are in the \code{Candidate} location, then uses a \PartitionCons primitive to move to the second phase where all processes are in locations \code{Leader} and \code{Replica} where the system is ready to receive requests from the clients.\ourskip

\para{Practical Parameterized Verification.}
Unfortunately, the decision procedure for PMCP in the \gspp fragment 
 has non-primitive recursive complexity~\cite{SchmitzS13}. 
Hence, we identify additional syntactic conditions, called {\em \amenability conditions}, that, for a given class of safety properties, enable reducing the parameterized verification problem for systems with \chwellbehaved processes to 
verification of a system with a small, fixed number of processes. This small, fixed number of processes 
is called a cutoff, and essentially entails a small model property: if there exists a counterexample to a safety property in a system with a certain, possibly large, number of processes, then there exists a counterexample to the property in a system with a cutoff number of processes.

\noindent{\em Example.} The cutoff for Distributed Store (\figref{casestudycode}) and its safety properties is 3. Essentially, due to the nature of the safety properties and the structure of the system, any violation of the property in a system with more than 3 actors can still be reproduced in a system with 3 process---any additional actors will not prevent the 3 process from potentially reaching an error state.
%

While the cutoff may seem obvious here, in general, cutoff arguments are non-trivial and deriving cutoff results requires a deep understanding of the underlying machinery for decidable parameterized verification.

We note that some \aml programs may not be practically verifiable. In this event, \kinarach provides best-effort feedback suggesting program modifications to the designer that can help {\em fit} their design into the desirable fragment of \aml programs.

\section{The \aml Modeling Language}
\label{sec:prelim} 

We present \aml, a language for modeling  distributed
agreement-based systems.
 We focus on systems in which the uncertainties and intricacies of their behavior in the presence of asynchronous communication and failures are essentially encapsulated within the underlying agreement protocols. Thus, while the agreement protocols may use asynchronous communication and tolerate network and process failures, we impose some simplifying assumptions on the system model {\em outside of} agreement. We assume that non-communicating processes can operate asynchronously, but communication is synchronous (\ie, sending and receiving processes must block until they can communicate). We further assume that processes may crash (i.e. exhibit \failstop failures), but the network is reliable. These assumptions enable our initial exploration of the boundaries of decidable parameterized reasoning for agreement-based systems. 

 While \aml includes standard features like communication actions, events, and event handlers, its distinguishing feature is the availability of {\em special primitives for encapsulating different agreement protocols}. 
\aml also includes some design choices to
facilitate decidable parameterized verification. We define the syntax and
semantics of \aml programs, with a detailed treatment of the semantics of its
agreement primitives.

%

\subsection{\aml Syntax and Informal Semantics}
\label{sec:amlsyntax}

\para{Programs.} A \aml program is a collection of an unbounded number $n$ of {\em identical\footnote{This can be relaxed to allow a finite number of distinct process definitions.}
  system processes} $P_1, P_2, \ldots, P_n$ and an {\em environment process} $E$, 
communicating via events. Each system process 
has a unique process index (\pid) drawn from
the set $\Ind = \{1,2,\ldots,\n\}$.
\ourskip

\para{Processes.}
The syntax for a \aml system process is shown in \figref{dsl}. 
A process definition begins with a declaration of typed \terma{variables} and communication \terma{actions}, and is followed by a sequence
of locations wit aa designated \terma{initial} location.
The variable type \terma{idSet}
corresponds to sets of process indices; the domain of this type is unbounded as
the number of system processes is, in general, unbounded.  The variable type
\terma{int} \neww{has a fixed, finite range that is specified in the declaration}
---this is one of the
inbuilt restrictions in \aml to facilitate automated parameterized
verification. Communication actions either represent communication between
system processes or communication between the environment process and system
processes; further, communication actions are either broadcast actions (denoted
\terma{br}) involving communication from one process to all other processes or
rendezvous actions (denoted \terma{rz}) involving
communication between a pair of processes. Each communication action \actname\xspace
has an optional (finite) integer-valued {\em payload} field  
that can be retrieved via the expression \actname.\terma{\pld}.

Each location contains a set of event handlers that consists of an event and a
reaction to that event. An event can be the empty event (\entry\xspace), a
receive of a communication action (\terma{recv}), or one of two \aprimitives. A
handler for an empty event corresponds to a {\em non-reactive} action a process
may initiate, i.e., an internal computation or a send of a communication action. 
A \terma{\PartitionCons} primitive \consnamep\xspace 
has two parameters: the set of participants
and the number of {\em winners} 
to be chosen. The set of winners (resp. {\em losers}) is retrieved via the expression \consnamep\terma{.winS} (resp.
\consnamep\terma{.loseS}). A \terma{\ValueCons} primitive \consnamev\xspace has three
parameters: the set of participants, the number of proposals to be chosen, and
an optional variable, \nont{\optintvar}, from which a process proposes its value. 
The $k^{\text{th}}$ value from the set of decided values, \consnamev\terma{.\decVal}, is retrieved via the expression \consnamev\terma{.\decVal}\terma{[}$k$\terma{]}.

\begin{figure}
\begin{minipage}{\linewidth}
\footnotesize

\begin{tcolorbox}
\setlength{\grammarparsep}{5pt plus 1pt minus 1pt} 
\setlength{\grammarindent}{7.4em} 

\begin{grammar}
 <process> ::= `\machine' "proc" `;' `\dslvariables' <vars> `; '  `\dslactions' <acts> `; initial' <locs> 

%

<vars> ::= $\epsilon$ | <vars> `;' <vars>
	\alt `idSet' "idSetV" 												\hfill {\footnotesize Set of \pids initialized to `Empty'}
	\alt `int' `['"intConst"`,' "intConst"`]' "intV" `\assign' "intConst"		\hfill {\footnotesize User-initialized bounded integer}

<acts> ::=  <sysActs> `;' <envActs>

<sysActs> ::= $\epsilon$ | <sysActs>  `;' <sysActs>
	\alt `br' "\actname" `:' <\ad>										\hfill {\footnotesize Broadcast action}
	\alt `rz' "\actname" `:' <\ad>										\hfill {\footnotesize \Pairwise action}

<envActs> ::=  $\epsilon$ | `env' <sysActs>

<\ad> ::=  `unit' 														\hfill {\footnotesize Empty payload}
	\alt `int' `['"intConst"`,' "intConst"`]'									\hfill {\footnotesize Bounded integer payload}
\end{grammar}
\tcblower
\setlength{\grammarparsep}{5pt plus 1pt minus 1pt} 
\setlength{\grammarindent}{7.4em} 
\begin{grammar}

<locs> ::=  <locs> `;' <locs>
	\alt `location' "\locname" <handlers>

<handlers> ::=  $\epsilon$ | <handlers> `;' <handlers>
	\alt `on' <event> <reaction>

<event> ::=  _  																			\hfill {\footnotesize Empty event}
	\alt `recv('"\actname"`)' 																\hfill {\footnotesize Receive event} 
	\alt `Partition<'"\consnamep"`>('<idSetExp>`,' "intConst"`)' 							\hfill {\footnotesize \PartitionCons event}
	\alt `\ValueCons<'"\consnamev"`>'`('<idSetExp>`,' "intConst"`,' <\optintvar>`)'  				\hfill {\footnotesize  \ValueCons event}

<reaction> ::=  `do'  <stmt> 																\hfill {\footnotesize Unguarded reaction}
	\alt  `where('<bExp>`)' `do' <stmt> 													\hfill {\footnotesize Guarded reaction} 	
	\alt `win:' <stmt> `lose:' <stmt> 														\hfill {\footnotesize \PartitionCons reaction}

<stmt> ::= <stmt> `;' <stmt>
	\alt <updateStmt>
	\alt <\dslcommStmt> 
	\alt <controlStmt>

<updateStmt> ::= 
	 "intV" := <intExp>
	\alt "idSetV"`.add('<idExp>`)'
	\alt "idSetV"`.remove('<idExp>`)'


<\dslcommStmt> ::= `\dslsend('"\actname"`,' <\optintvar>`,' <idExp>`)'  							\hfill {\footnotesize \Pairwise send statement}
	\alt `\dslbroadcast('"\actname"`,' <\optintvar>`)' 											\hfill {\footnotesize Broadcast send statement}

<controlStmt> ::= `if('<bExp>`)' <stmt> `else' <stmt>
	\alt `goto' "\locname"

<\optintvar> ::= _` '| "intV"  																\hfill {\footnotesize \neww{Optional integer variable}}

\end{grammar}
\end{tcolorbox}
\caption{Syntax for \aml. In the grammar,
non-terminals are enclosed in \nont{ }, keywords are in \terma{boldface}, and all other terminals are \termb{monospaced}.
}
\label{fig:dsl}

\end{minipage}
\end{figure}

\begin{figure}
\footnotesize
\begin{tcolorbox}

\setlength{\grammarparsep}{6pt plus 1pt minus 1pt} 
\setlength{\grammarindent}{7em} 

\begin{grammar}

<idExp> ::= `self' 				\hfill {\footnotesize \Pid of current process}
	\alt "\actname"`.sID' 			\hfill {\footnotesize \Pid of sender of action "\actname"}

<idSetExp> ::= `All' | `Empty' | "idSetV" 
	\alt "\consnamep"`.winS' 		\hfill  {\footnotesize Set of winners of \PartitionCons primitive "\consnamep"}
	\alt "\consnamep"`.loseS'		\hfill  {\footnotesize Set of losers of \PartitionCons primitive "\consnamep"}

<intExp> ::= "intConst" | "intV" | <intExp> <arithOp> <intExp>
	\alt  "\actname"`.\pld'  		\hfill{ \footnotesize Payload of action "\actname"}
	\alt "\consnamev"`.\decVal'`['"intConst"`]' 		\hfill{ \footnotesize Selecting some decided value of \ValueCons primitive "\consnamev"}



<bExp> ::= `True' | `False' | `!'<bExp>
	\alt <bExp>  <boolOp>  <bExp>
	\alt <intExp>  <cmpOp>  <intExp>
	\alt <idExp>  <eqOp>  <idExp>

<cmpOp> ::= `<' | `>' | `<=' | `>=' | <eqOp>

<eqOp> ::= `=' | `!='

\end{grammar}
\end{tcolorbox}
\caption{Syntax of \aml Expressions.}
\label{fig:dslexp}
\end{figure}

\begin{figure}
\footnotesize
\begin{tcolorbox}

\setlength{\grammarparsep}{6pt plus 1pt minus 1pt} 
\setlength{\grammarindent}{7em} 

\begin{grammar} 
%
%

<handlers> ::= `passive' <eventList> \hfill {\footnotesize Handler specifying list of events a process should {\em not} react to}

<eventList> ::= <eventId> | <eventId> `,' <eventList> 

<eventId> ::= "\actname" | "\consnamep" | "\consnamev"
 
<reaction> ::=  `reply('"\actname"`,' <\optintvar>`)' \hfill \ {\footnotesize A \pairwise~{\em reply} to the last sender}
\end{grammar}
\end{tcolorbox}
\caption{\aml Syntactic Sugar. 
}
\label{fig:dslsugar}
\end{figure}

A reaction to an event consists of a block of update statements,
control statements, and/or sends.  
Update statements include assignments to integer variables 
and statements to add or remove a \pid from a set of \pids.
Control statements include
conditionals and \terma{goto} statements used to switch between locations.  A
\terma{\dslsend} statement is a rendezvous send that transmits a message with
an action identifier \actname~and an optional payload \nont{\optintvar} to a process with index
given by \nont{idExp}. A \terma{\dslbroadcast} statement is a broadcast send
that transmits a message with an action identifier \actname~and an optional
payload \nont{\optintvar} to all other processes.  

Reactions for empty, receive and \code{\ValueCons} events all begin with \terma{do}. 
Additionally, a \emph{guarded reaction} of the form \terma{where(}\nont{bExp}\terma{)} \terma{do} \nont{stmt} can be used for empty and receive events to ensure that the handler is only enabled if some Boolean predicate evaluates to \terma{true}.
Finally, the reaction for a \terma{\PartitionCons} event is of the form \terma{win:}
\nont{stmt} \terma{lose:} \nont{stmt}, indicating how a process should react if
it wins and if it loses.

An environment process is a simpler version of a system process with variable
types restricted to \terma{int} and event types restricted to 
empty and receive events.\ourskip
%

\para{Expressions.}  The syntax for all expressions in \aml processes is
shown in \figref{dslexp}. 
An \nont{idExp} expression evaluates to a \pid---\terma{self} 
retrieves the \pid of the current process and, in a receive handler for communication action $\actname$, the expression
\actname\terma{.sID} retrieves the \pid of the corresponding sender.  
An \nont{idSetExp} expression evaluates to a set of \pids as shown
and includes the expressions \consnamep\terma{.winS} and \consnamep\terma{.loseS}
introduced earlier. An \nont{intExp} expression evaluates to an
integer and includes the expressions \actname.\terma{\pld} and
\consnamev\terma{.\decVal}\terma{[}\termb{intConst}\terma{]}, introduced earlier. 
A \aml arithmetic expression, \nont{arithOp}, 
is standard and is not shown. A \aml Boolean expression, \nont{bExp}, 
constrains comparison of \nont{idExp} expressions to equality and disequality
checks; we expand on this restriction at the end of \secref{amlbasic} and 
emphasize that 
this is a common syntactic restriction used to facilitate the use of
{\em structural symmetries} for scalable verification (cf. \cite{emerson1996symmetry,emerson2003combining,wahl2007adaptive,Gleissenthall.Pretend.Synchrony.POPL.2019,ip1996better}).\ourskip

\para{Syntactic Sugar.} \aml provides syntactic sugar (\figref{dslsugar}) to simplify expressing some common idioms. 
A \terma{passive} handler specifies events a process should {\em not} react to. A \terma{reply} reaction sends a \pairwise reply to the last sender.

\subsection{Agreement-Free \aml Program Semantics}
\label{sec:amlbasic} 

The semantics of \aml processes and programs is best described using 
state-transition systems. We first define the semantics of \aml programs 
without \aprimitives, then extend the definition to \aml programs with 
\aprimitives in \secref{choosemodel}. Intuitively, the semantics 
allows non-communicating processes in \aml programs to operate
asynchronously while ensuring that communication and agreement is synchronous and consistent.\ourskip

\begin{figure}[thp]
\begin{minipage}{0.85\linewidth}
\footnotesize
\begin{tcolorbox}

\setlength{\grammarparsep}{6pt plus 1pt minus 1pt} 
\setlength{\grammarindent}{4em} 
\begin{grammar}

<iStmt> ::=  <uStmts>`;' `goto' "\locname"

<sStmt> ::= <\dslcommStmt>`;' <iStmt>

<uStmts> ::=  $\epsilon$ | <updateStmt> | <uStmts>`;' <uStmts> 

\end{grammar}
\end{tcolorbox}
\end{minipage}
\begin{subfigure}{0.1\linewidth}
\caption{}
\label{fig:corestmts}
\end{subfigure}

\begin{minipage}{0.85\linewidth}
\footnotesize
\begin{tcolorbox}[colback=white,sharp corners,boxrule=0.3mm,right=0.0mm, top=-1.5mm,bottom=-1.5mm]
\begin{udsl}
location (*\locname*)
 on _ where ((*\nontDSL{bExp}*)) do
  (*\nontDSL{iStmt}*)
\end{udsl}
\end{tcolorbox}
\end{minipage}
\begin{subfigure}{0.1\linewidth}
\caption{} 
\label{fig:handlera} 
\end{subfigure}
\begin{minipage}{0.85\linewidth}
\footnotesize
\begin{tcolorbox}[colback=white,sharp corners,boxrule=0.3mm,right=0.0mm, top=-1.5mm,bottom=-1.5mm]
\begin{udsl}
location (*\locname*)
 on _ where ((*\nontDSL{bExp}*)) do
  (*\nontDSL{sStmt}*)
\end{udsl}
\end{tcolorbox}
\end{minipage}
\begin{subfigure}{0.1\linewidth}
\caption{} 
\label{fig:handlerb} 
\end{subfigure}

\begin{minipage}{0.85\linewidth}
\footnotesize
\begin{tcolorbox}[colback=white,sharp corners,boxrule=0.3mm,right=0.0mm, top=-1.5mm,bottom=-1.5mm]
\begin{udsl}
location (*\locname*)
 on recv((*\actname*)) where ((*\nontDSL{bExp}*)) do
  (*\nontDSL{iStmt}*)
\end{udsl}
\end{tcolorbox}
\end{minipage}
\begin{subfigure}{0.1\linewidth}
\caption{} 
\label{fig:handlerc}
\end{subfigure}

\begin{minipage}{0.85\linewidth}
\footnotesize
\begin{tcolorbox}[colback=white,sharp corners,boxrule=0.3mm,right=0.0mm, top=-1.5mm,bottom=-1.5mm]
\begin{udsl}
location (*\locname*)
 on (*\terma{\PartitionCons}*)<(*\consnamep*)>((*\nontDSL{idSetExp}*),(*\termb{intConst}*)) 
  win: goto (*\locname*) lose:  goto (*\locname*)
\end{udsl}
\end{tcolorbox}
\end{minipage}
\begin{subfigure}{0.1\linewidth}
\caption{} 
\label{fig:handlerd} 
\end{subfigure}

\begin{minipage}{0.85\linewidth}
\footnotesize
\begin{tcolorbox}[colback=white,sharp corners,boxrule=0.3mm,right=0.0mm, top=-1.5mm,bottom=-1.5mm]
\begin{udsl}
location (*\locname*)
 on (*\terma{\ValueCons}*)<(*\consnamev*)>((*\nontDSL{idSetExp}*),(*\termb{intConst}*),(*\nontDSL{\optintvar}*))
   goto (*\locname*)
\end{udsl}
\end{tcolorbox}
\end{minipage}
\begin{subfigure}{0.1\linewidth}
\caption{} 
\label{fig:handlere} 
\end{subfigure}

\caption{Syntax of Core \aml. (a) Core Statements. Core handler for (b) Internal, (c) Send, (d) Receive, (e) \PartitionCons, and (f) \ValueCons.}
\label{fig:handler-fig}
\end{figure}

\para{Core Fragment of \aml.}
To enable a succinct description of \aml programs' semantics, 
we rewrite process definitions
into a core fragment of the language with the event handlers and statements 
depicted in \figref{handler-fig}. The handlers may contain two types of statements (shown in \figref{corestmts}): a statement \nont{iStmt} that consists of a (possibly empty) sequence of update statements, followed by a \terma{goto} statement; and a statement \nont{sStmt} that consists of a send statement, followed by an \nont{iStmt} statement.
The \emph{internal} core handler in \figref{handlera} embodies computations that the process does 
without communication with other processes; 
the \emph{send} core handler in \figref{handlerb} embodies a send of some action by a process; and the \emph{receive} core handler in \figref{handlerc} embodies the reaction of a process to a receive of some action. Note that all three handlers are guarded by a predicate that dictates when they are enabled. The core handlers in \figref{handlerd} and \figref{handlere} are for \aprimitives and only contain \terma{goto} statements as shown. 
For the rest of this paper, let $P$ be a process in the core frament of \aml. 
With some abuse of notation, we use $\V$, $\acts$, and $\Q$ to refer to the sets of variables,
actions, and locations in their eponymous sequences in~\figref{dsl}. 
\ourskip

\para{Process Semantics.} The semantics of a process $P$ is defined as a 
labeled state-transition system 
$(\LQ,\loq_0,\crashstate,\acts,\T)$, where $\LQ$ is the set of (local) states, $\loq_0$ is the
initial state, $\crashstate$ is a special ``crashed'' state, $\acts$ is the set of 
actions, and $\T \subseteq \LQ \times \{ \mathtt{sendrz},\mathtt{sendbr},$
$\mathtt{recvrz},\mathtt{recvbr},\crash,\epsilon\} \times \acts \times \Ind \times \LQ$
is the set of (local) labeled transitions of $P$.
A state $\loq \in \LQ$ is a pair $(\q,\sigma)$ where $\q \in \Q$ is a location and $\sigma$ is a valuation of the variables in $\V$. We let $\sigmaOf{var}$ denote the value of the variable $var$ according to $\sigma$.
For a state $\loq = (\q,\sigma)$, we let $\loq.loc$ denote the location $\q$ in $\loq$, and 
$ \valOf{\loq}{var}$ denote the value $\sigmaOf{var}$ of variable $var$ in $\loq$. 
Similarly, we use $\sigmaOf{expr}$ ($\valOf{\loq}{expr}$) to denote the value of 
expression $expr$ evaluated under $\sigma$ (in state $\loq$).
The initial state $\loq_0 = (\q_0, \sigma_0)$, where $\q_0$ denotes the initial
location and $\sigma_0$ denotes the initial variable valuation. The crash state $\crashstate$ is a special state that the process is assumed to enter upon  exhibiting a \failstop failure.

A transition of process $P$ without \aprimitives corresponds 
to the execution of one of the three core event handlers in \figref{handlera}, \figref{handlerb}, and \figref{handlerc}; a transition 
is labeled either with a send/receive of a communication 
action in $acts$ or an empty label $\epsilon$ denoting an internal transition.
For each core handler, let $\q$ denote the current location and $\q'$ denote the target location of the \terma{goto} statement. 
Then, the transitions in $T$ are defined as follows: 
\begin{compactenum}[(a)]
%

\item For each broadcast send handler as shown in \figref{handlerb}
  with \nont{sendStmt} given by
  \terma{\dslbroadcast}(\actname, \nont{\optintvar}), $T$ contains a transition
  $\sendbr{(\q,\sigma)}{\actname}{(\q',\sigma')}$ for each $\sigma$ such
  that $\sigma(bExp)= true$ and $\sigma'$ is obtained from $\sigma$ by
  applying the sequence of updates \nont{uStmts}. Note that if
  \nont{\optintvar} is $\epsilon$, the payload is empty and if
  \nont{\optintvar} is a variable, denoted by $var_{\actname}$, the payload is
  $\sigmaOf{var_\actname}$.


\item For each broadcast receive handler as shown in
  \figref{handlerc} with broadcast action \actname, $T$ contains a transition
  $\recvbr{(\q,\sigma)}{\actname}{(\q',\sigma')}$ for each $\sigma$ such
  that $\sigma(bExp) = true$ and $\sigma'$ is obtained from $\sigma$ by
  applying the sequence of updates \nont{uStmts}. Note that \nont{uStmts} may access the received value using the expression $\actname.\terma{\pld}$.

\item To model process \failstop failures, $T$ contains a transition
  $(\q,\sigma)\xrightarrow{\crash}\crashstate$ for each ($\q,\sigma$).

\end{compactenum}

%



Local transitions corresponding to internal and \pairwise send and receive can be formalized similarly. We use $E$ to denote the environment process and $\LQ_E, \loq_{0,E}$ etc. 
to denote its set of states, initial state etc., respectively.\ourskip

\para{Distributed Program Semantics.}
The semantics of a \aml program consisting of 
$\n$ identical system processes $P_1, \ldots, P_\n$ and the
environment process $E$ 
is defined as a state-transition system $\M(n) =
(\GQ,\gq_0,\GT)$, parameterized by the number of processes $n$, 
where:
\begin{compactenum} [1.]
\item $\GQ = \LQ^\n \times \LQ_E$ is the set of global states, 
\item $\gq_0 = (\loq_{0},\ldots,\loq_{0},\loq_{0,E})$ is the initial global state, and
\item $\GT \subseteq \GQ \times \GQ$ is the set of global transitions
%
%
%
%
where (i) 
all processes synchronize on a broadcast communication action, (ii) two processes synchronize on a rendezvous communication action, (iii) one process makes an asynchronous internal move, or (iv) one process crashes. Formally, a global transition $(\gq, \gq')$ based on a broadcast  action $\actname$ is in $\GT$ iff  there exists a process $P_i$ with  
  $\sendbr{\loq_i}{\actname}{\loq'_i}$, and 
  every other process $P_j$ with $j \neq i$ has a 
  transition
  $\recvbr{\loq_j}{\actname}{\loq'_j}$ such that 
	$\gq'=\gq[\loq_i \gets \loq_i', \forall j \neq i: \loq_j \gets \loq_j']$ 
	and if \nont{\optintvar} is the variable $var_\actname$,
	$\valOf{\loq_i}{var_\actname} = \actname.\terma{\pld}$.
	Here, $\gq[\loq_i \gets \loq_i']$ indicates that process $P_i$ moves from state $\loq_i$ to $\loq_i'$. 
	A global crash transition $(\gq, \gq')$ is in $\GT$ iff there exists a process $P_i$ with a local crash transition
	   $\loq_i$ $\xrightarrow{\text{crash}}$ $\crashstate$ and $\gq'=\gq[\loq_i \gets \crashstate]$.
	Global transitions corresponding to rendezvous actions or internal transitions can be formalized similarly.
	%
\end{compactenum}
\ourskip



An execution of a global transition system $\M(n)$ is a (possibly infinite) 
sequence of states, $\gq_0, \gq_1,\ldots$, in $\GQ$ such that for each $j \geq 0$, 
$(\gq_j,\gq_{j+1}) \in \GT$. 
A state $\gq$ is {\em reachable} if there exists
a finite execution of $\M(n)$ that ends in $\gq$.

In what follows, we use $\M$ and $\M(n)$ as well as 
system process and process, interchangeably. 
\ourskip

\para{Correctness Specifications.}
In this work, we focus on a broad class of invariant properties of systems
modeled in \aml.  In particular, our correctness specifications are Boolean
combinations of universally quantified formulas over locations, \terma{int} variables, and a finite number of variables with distinct valuations over $\Ind$.


For example, one can specify that a location $c$ is a critical section (of size
1) as: $\forall i, j \in \Ind.\ \neg(q[i].loc = c \land q[j].loc = c)$; Distributed Store
(\figref{casestudycode}) uses a specification of this form to ensure that at most 1 process can be in \code{Leader}. As another
example, one can specify that all processes in some location $d$ must have the
same value in their local variable $v$ as: $\forall i,j \in \ \Ind. q[i].loc = d
\land q[j].loc = d  \Rightarrow \valOf{q[i]}{v} = \valOf{q[j]}{v}$; Distributed Store uses specifications of this form to ensure the stored data is consistent.



The program $\M(n)$ is {\em safe} if it has no reachable states that violate its
correctness specification. Given a specification $\phi(n)$, also parameterized by $n$, we use the
standard notation $\M(n) \models \phi(n)$ to denote that $\M(n)$ is safe.\ourskip

\para{Symmetry for Efficient, Parameterized Verification.} 
Performing automated parameterized verification for systems with an
arbitrary number of processes hinges on the number of different {\em types} of
processes being {\em bounded} (in \aml, there are two types: system and environment). Thus, parameterized systems naturally
exhibit many {\em similar} global behaviors that are independent of specific
process indices. The symmetric nature of such global behaviors
offers another advantage: it is possible to greatly improve
the verification time of symmetric systems through 
{\em symmetry reduction}~\cite{emerson1996symmetry}. In particular, a (global) state-transition
system $\M$ is \emph{fully-symmetric} if its transition relation
$\GT$ is invariant under permutations over the set $\Ind$ of \pids.
As noted in \secref{amlsyntax}, \aml syntactically constrains comparison of
\nont{idExp} expressions 
to (dis)equality checks. 
This is a sufficient condition to ensure \aml processes are fully-symmetric 
and, hence, enable parameterized verification and symmetry reduction. 

\subsection{Semantics of \aml Agreement Primitives}
\label{sec:choosemodel}

We now extend the process and program semantics defined in
\secref{amlbasic} to \aml programs with \aprimitives. Furthermore, we show that
our definition of the semantics of agreement primitives provides a sound
abstraction of agreement protocols and enables symmetry
reduction. 

To simplify the presentation of the semantics, we expand the set $\V$ of
variables as follows.  For each \PartitionCons event \consnamep, we add
variables \consnamep\termb{_winS} and \consnamep\termb{\_loseS} for storing the
sets of winners and losers, respectively.  Similarly, for each \ValueCons event
\consnamev, we add variable \consnamev\termb{_\decVal} for storing the decided
values.\ourskip

\para{Process-level Semantics of \Aprimitives.} For the  \PartitionCons event
handler (\figref{handlerd}), let $\q$ be the current location and $\q_w$
(resp. $\q_l$) be the target location of the \code{goto} statement in the
\terma{win:} (resp. \terma{lose:}) block.  For the \ValueCons event handler
(\figref{handlere}), let $\q$ denote the current location and $\q_d$ denote the
target location of the \code{goto} statement.  Then, the set $T$ of transitions
is extended as follows: 

\begin{compactenum}[(a)]
\item For each \PartitionCons handler with event
  \terma{\PartitionCons} <\consnamep> (\nont{idSetExp},\termb{intConst}) 
  in \figref{handlerd}, $T$ contains transitions
  $(\q,\sigma)$ $\xrightarrow{\pcw(\chset,\chcard)}$ $(\q_w,\sigma')$ and
  $(\q,\sigma)$ \\ $\xrightarrow{\pcl(\chset,\chcard)}$ $(\q_l,\sigma')$ for each $\sigma$ such that
  $\chset$, matched by \nont{idSetExp}, denotes the set of participants\footnote{While such sets are usually
    predefined, we allow more flexibility by permitting processes to
    communicate and construct them.},
  $k$, given by \termb{intConst}, denotes
  the number of winners to be decided, and
  $\sigma'$ is obtained from $\sigma$ by updating variables
  \consnamep\termb{_WinS }and \consnamep\termb{_loseS} to the sets of winners
  and losers in the global invocation of \consnamep, respectively. 
  
\item For each \ValueCons handler with event
  \terma{\ValueCons} <\consnamev> (\nont{idSetExp}, \termb{intConst},
  \nont{\optintvar}) in \figref{handlere}, $T$ contains a local
  transition $(\q,\sigma)$ $\xrightarrow{ \vc(\chset,\chcard,\chvar)}$ $(\q_d,\sigma')$ for each $\sigma$ 
  such that $\chset$ and $k$ are as before, $\chvar$ denotes the variable
  from which a process proposes its value, matched by \nont{\optintvar}
  if \nont{\optintvar} is not $\epsilon$, and $\sigma'$ is obtained from
  $\sigma$ by updating the variable \consnamev\termb{_\decVal} to the decided
  values in the invocation of \consnamev.\ourskip
\end{compactenum}

\para{Program-level Semantics of \Aprimitives.} The local transitions
corresponding to \aprimitives are essentially {\em modeling} invocation of
verified \consagree protocols that enable a set of participants to decide on a
finite set of winners/values in a {\em globally consistent} way. As stated in
\secref{overview}, verified agreement protocols typically entail agreement,
validity, and termination.  Thus, to ensure that agreement primitives provide a
sound abstraction of verified agreement protocols, the \emph{global behavior}
of these primitives must satisfy a set of conditions entailed by agreement,
validity, and termination. 
We represent this set of conditions on the global transitions corresponding to
\aprimitives as a precondition-postcondition pair, stated informally as:
\begin{compactenum}

\item [$C_1$:] {\em Consistent Participants Precondition}. The participants agree on
  {\em with whom} to invoke agreement
  \footnote{
    Systems in which
    \emph{all} processes intend to reach agreement trivially satisfy
    $C_1$. The more general form of $C_1$ enables systems to
    invoke agreement protocols with only a subset of processes. 
    }
    , and,

\item [$C_2$:] {\em Consistent Decisions Postcondition}.  Upon termination of agreement, 
  all \emph{\noncrashed} participants concur on winners/values.

\end{compactenum}

In what follows, we present the global transitions and specialization of the 
precondition-postcondition pair (C1,C2) for each type of \aprimitive.\ourskip

\noindent{\em \PartitionCons.} Consider an instance of a
\PartitionCons \aprimitive with identifier \consnamep~and local transitions
$(\q^c,\sigma)$ $\xrightarrow{\pcw(\chset,\chcard)}$ $(\q^c_w,\sigma')$ and
$(\q^c,\sigma)$ $\xrightarrow{\pcl(\chset,\chcard)}$ $(\q^c_l,\sigma')$. 
Let $\chlocsp$ be the set of all locations $\q^c$ from which the
participants of this instance may invoke \PartitionCons (i.e., all locations where
the above two transitions originate). 

We extend the global transition relation $\GT$ of $\M$ with 
a \PartitionCons \consagree transition from global state $\gqs$ to
global state $\gqe^\winset$ encoding a selected set $\winset$ 
of $\chcard$\footnote{
All \noncrashed participants act as winners if the number of \noncrashed participants is less than $k$.
} \noncrashed winners, and a set $\failset$ of participants that have crashed during agreement if:

\begin{itemize}

  \item [$C_1(PC) $:] There exists a set $\gchset \subseteq \Ind$ of processes in $\gqs$ in 
    appropriate locations for invoking this instance of the \PartitionCons primitive 
    and with a consistent view of each other. Formally:
	
  \begin{compactenum}
	\item $\forall i \in \gchset : {\gqs}[i].loc \in \chlocsp$ and 
	\item $\forall i,j \in \gchset \!:\! \valOf{\gqs[i]}{\chset} \!=\! \valOf{\gqs[j]}{\chset} \!= \! \gchset$, and,

	\end{compactenum}

\item [$C_2(PC)$:] The \noncrashed processes of $\gchset$
  move to their appropriate target locations in $\gqe^\winset$ based on whether
  they {\em win} or {\em lose}  and their \consnamep\termb{_winS} and
  \consnamep\termb{_LoseS} variables in $\gqe^\winset$ are updated to reflect
  the partition while the set $\failset \subset \gchset$ of crashed processes move to the crash state $\crashstate$. As explained in Remark 1 below, we assume that if all the participants fail, then no valid $\gqe^\winset$ exists. Formally:
	Let $N$ be the set $\gchset \setminus \failset$ of \noncrashed participants, then:
  \begin{compactenum} 
	 \item $\forall i  \in N :  i \in \winset  \wedge \gqs[i].loc  =  \q^c  \Rightarrow  \gqe^\winset[i].loc =  \q^c_w$,
	 \item $\forall i  \in N:  i \not \in \winset  \wedge \gqs[i].loc = \q^c  \Rightarrow  \gqe^\winset[i].loc = \q^c_l$,
	\item $\forall i  \in N  : \valOf{\gqe^\winset[i]}{\consnamep\termb{\_winS}} = \winset$,
	\item $\forall i  \in N  : \valOf{\gqe^\winset[i]}{\consnamep\termb{\_LoseS}} = N \setminus \winset$,
	\item $\forall i  \in \failset : \gqe^\winset[i] = \crashstate$, and,
	\item $\forall i  \in \Ind \setminus \gchset  : \gqe^\winset[i] = \gqs[i]$.
	\ourskip
%
%
%
	\end{compactenum}
\end{itemize}

\noindent{\em \ValueCons.} Consider an instance of a \ValueCons agreement
primitive with identifier \consnamev~and local transition
$(\q^c,\sigma)$ $\xrightarrow{\vc(\chset,\chcard,\chvar)}$ $(\q^c_d,\sigma')$. As
before, let $\chlocsv$ be the set of locations $\q^c$ from which the
participants of this \consnamev~instance may start.  

We extend the global transition relation $\GT$ of $\M$ with 
a \ValueCons
\consagree transition from a global state $\gqs$ 
to a global state $\gqe^\winset$ encoding a selected set $\winset$ of $\chcard$ decided
values\footnote{Note that, a proposed value of a crashed process can still be chosen as the decided value.} and a set $\failset$ of participants that have crashed during agreement if:

\begin{itemize}
  \item [$C_1(VC)$:] The state $\gqs$ is as defined for \PartitionCons, and, 
  \item [$C_2(VC)$:] The set $N \coloneq \gchset \setminus \failset$ such that $|N|>|\failset|$ of \noncrashed processes move to their target locations in $\gqe^\winset$ 
    and their \consnamev\termb{_\decVal} variables are updated to reflect the decided values while the set
    $\failset$ of crashed processes move to the crash state $\crashstate$.
    As explained in Remark 1 below, we assume that if a majority of participants fail (i.e., $|N|\leq|\failset|$), then no valid $\gqe^\winset$ exists. Formally:    

\begin{compactenum}
\item $\forall i \in N: \gqs[i].loc = \q^c \Rightarrow
  \gqe^\winset[i].loc = \q^c_d$, 
\item $\forall i \in N: \;
  \valOf{\gqe^\winset[i]}{\consnamev\termb{\_\decVal}} = \winset$, 
\item $\forall i  \in \failset : \gqe^\winset[i] = \crashstate$, and,  
\item $\forall i \in \Ind \setminus \gchset: \gqe^\winset[i] = \gqs[i]$.
\end{compactenum}
\end{itemize}
\ourskip

\noindent{\em Remark 1: Failure Assumptions for Valid Termination.}
 Common agreement protocols have assumptions about process failures under which they guarantee the validity of results upon termination. For instance, leader election protocols~\cite{bullyalgo} require the elected leader to not fail, but can tolerate the failures of the losing processes and consensus protocols like Paxos~\cite{lamport1998part} and Raft~\cite{ongaro2014search} require a simple majority of the participants to not fail and to agree on a proposed value. 
	While, in general, the semantics of \aml's agreement primitives can be parameterized over specific failure assumptions, our default definitions
	 encode these common assumptions. Thus, when using the \PartitionCons primitive, any global state $\gqe^\winset$ where all the participants have failed is assumed to be not valid. When using the \ValueCons primitive, any global state $\gqe^\winset$ where a majority of the participants have failed is assumed to be not valid.
\ourskip

\para{Soundness.}
\aml's agreement primitives are sound: 

\begin{lemma}
\label{lem:abs-correctness}
Our proposed abstraction of verified agreement protocols, as defined using the syntax and semantics of \aml \aprimitives, is sound. \neww{In other words, if an agreement protocol satisfies agreement, validity, and termination, then the agreement protocol satisfies the semantics of \aprimitives captured by the precondition-postcondition
pair $(C_1,C_2)$.}
\end{lemma}

\noindent{\em Proof.}
We prove this by contradiction. \neww{Assume that an agreement protocol 
  satisfying agreement, validity, and termination begins in a state $\gqs$ that satisfies precondition $C_1$\footnote{A violation of precondition $C_1$ (i.e., participants do not have a consistent view of each other or are in invalid local states to participate in the agreement protocol) indicates an invalid global state to invoke agreement.} but ends in a state $\gqe^\winset$ that violates postcondition
  $C_2$.}
  A violation of postcondition $C_2$ (i.e., participants not 
  agreeing on the same winner/value or agreeing on a winner/value that 
  was not in the set of participants/was never proposed)
  contradicts agreement and validity.
  Finally, a violation due to the absence of a transition between 
  a state $\gqs$ satisfying $C_1$ and a state $\gqe^\winset$ satisfying 
  $C_2$ directly contradicts termination.
  
Note that the statement of  \lemref{abs-correctness} implicitly assumes that the failure assumptions encoded in the semantics of \aml agreement primitives hold. 
  If the failure assumptions do not hold, neither our primitives nor the agreement protocols they abstract provide any guarantees.
  \ourskip

\para{Symmetry.}
In a state-transition system, $\M_{\ch} = (\GQ,q_0,\GT)$, capturing the semantics of a \aml program, 
let $\GT_{\ch}$ be the set of
all transitions corresponding to \aprimitives in $\GT$.  
Let $\M = (\GQ,q_0,\GT\setminus \GT_{\ch})$ be the state-transition system
without the {\em \consagree transitions} of $\M_{\ch}$.\footnote{We do not make any symmetry-related assumptions about the specific agreement protocol that an \aprimitive 
encapsulates. In particular, the underlying agreement protocol could employ 
non-symmetric strategies such as ``the process with maximum \pid wins''.}

\begin{lemma}
\label{lem:indexIndepOfCh}
If $\M$ is fully-symmetric, then $\M_{\ch}$ is fully-symmetric.
\end{lemma}

Intuitively, the proof (ref. \appref{chsymmetry}) is based on the observation that 
\consagree transitions are oblivious to the identities of the participants 
and are hence invariant under permutations over $\Ind$.  
\ourskip

\section{Verification of \aml Programs}
\label{sec:verification}

We now formalize the parameterized verification problem 
for \aml programs and present our theoretical results for enabling 
decidable and efficient parameterized verification.\ourskip

\para{\aml Parameterized Verification Problem (\mpvp).}
Given a \aml system process $P$, an environment process $E$, 
and a parameterized safety specification 
$\phi(n)$ as defined in \secref{prelim}, \mpvp asks if $\forall n. \; \M(n) \models \phi(n)$.

Our first result (\secref{decidablefragment})
  identifies conditions on \aml programs and the specification $\phi(n)$
  for enabling decidability of \mpvp. 
  Our second result (\secref{cutoffs}) identifies additional conditions for
  which this problem is {\em efficiently} decidable, based on {\em cutoff}
  results.  Cutoff results reduce the
  parameterized verification problem to a verification problem over a {\em
    fixed} number of processes.  Formally, 
a cutoff for a parameterized system $\M$ and correctness specification $\phi$ is 
a number $c \in \Nats$ such that:
\[ 
  \forall n \geq c. \; (\M(c) \models
    \phi(c) \iff \M(n) \models \phi(n)).
\] 
In particular, our second result identifies conditions on \aml programs for {\em small} cutoffs, reducing \mpvp to verification of a \aml program 
with a {\em small} number of processes. The latter problem is decidable for any 
\aml program, as the corresponding semantics can be expressed as a finite-state machine.

For the rest of this section, we fix a \aml process $P$ with a set of process-local states $\LQ$, initial state $\ls_0$, and process-local transitions $\T$, and refer to the corresponding global state-transition system, $\M$, as a (\aml) system.


\subsection{Decidable Parameterized Verification}
\label{sec:decidablefragment}
%

%
%

%
%


\newww{\para{Phases.}}  
To enable decidable parameterized verification, we view \aml systems as proceeding in \emph{\phases}. A phase of a \aml system is a set of process-local states, characterizing the set of events that can occur when {\em all} processes co-exist in that set of local states. In any global execution, all processes simultaneously move from one phase to the next, where a new set of events may occur. Processes move between phases strictly via \emph{globally-synchronizing events}, i.e, broadcasts or \aprimitives; within a phase, processes can use any type of communication. 
While two phases may share some local states, their associated events are disjoint. We note that any \aml system can be viewed as proceeding in phases, by identifying phases with appropriate sets of events---so phases do not constrain the applicability of our approach.\ourskip 
\begin{newtextt}

In what follows, 
we refer to the set of globally-synchronizing events 
 as $\gsactions$ and
 the set of rendezvous actions in $P$ as $\pwactions$. 
For each event $\act$, we define its source set, denoted $src_\act$,
as the set of states in $\LQ$ {\em from which} 
there exists a transition in $\T$ labeled with $\act$.
Similarly, we
define the destination set of each event $\act$, denoted $dst_\act$, 
as the set of states in $\LQ$ {\em to which} a transition in $\T$ labeled with $\act$ 
exists.
 For instance, if $\act$ is a broadcast action $\actname$, 
$src_\actname = \{\loq \mid \sendbr{\loq}{\actname}{\loq’} \in \T\ \vee \
\recvbr{\loq}{\actname}{\loq’} \in \T\}$ and 
$dst_\actname = \{\loq' \mid \sendbr{\loq}{\actname}{\loq’} \in \T \ \vee \
\recvbr{\loq}{\actname}{\loq’} \in \T\}$. 
The source and destination sets for rendezvous actions and 
instances of \ValueCons and \PartitionCons can be defined similarly. 
Finally, we define the relation $\relatedIntPw \subseteq \LQ \times \LQ$ to
denote pairs of states related via internal or \pairwise transitions as follows:
$$
\relatedIntPw = \{(s,t) \mid s \neq t \; \wedge \;   
(\trans{s}{\epsilon}{t} \in \T \, \vee \, \trans{t}{\epsilon}{s} \in \T  \, \vee \, 
  (\exists \act \in \pwactions.\ \{s, t\} \subseteq src_\act \cup dst_\act )) \}.
$$


We now present a constructive definition for 
the set of phases of $S$. Intuitively, two states are in the same phase if 
they
are part of the same source or destination set, or, their phases are 
{\em connected} by internal or \pairwise transitions.

\begin{definition}[Phases]
  \label{def:phases}
The set of phases is constructed as follows:  
  \begin{enumerate}
\item {\em Initialization}:
The set of phases is initialized to the 
set of source sets and destination sets of each globally-synchronizing event:
$$
\initPhases = \bigcup_{\act \in \gsactions} \big{\lbrace} src_\act,\ dst_\act \big{\rbrace}.
$$
Informally, the source set of a globally-synchronizing event $\act$ is a subset of the local state space where \emph{all} processes need to co-exist for $\act$ to occur. The destination set of $\act$ characterizes the set of states in which \emph{all} processes co-exist after event $\act$ has occurred.
\item {\em Expansion}:
Each initial phase is then expanded such that if a state $s$ is in a phase, then every  state $t$ such that $\relatedIntPw(s, t)$ holds is in the phase too:
$$
\expandedPhases = \bigcup_{X \in \initPhases} \big{\lbrace} X \cup \{ t \mid s \in X \wedge \relatedIntPw^+(\ls, t)\} \big{\rbrace},
$$
where $\relatedIntPw^+$ is the transitive closure of $\relatedIntPw$. Informally, this step ensures 
that any local state that is reachable from or can be reached by a state in an initial phase via internal or \pairwise transitions, is added to that phase.


\item {\em Merge}:
Finally, expanded phases that contain distinct states $s$, $t$ with $\relatedIntPw(s, t)$ are merged:
$$
\phases =\big{\lbrace} \bigcup_{W \in X} W \mid  X \subseteq \expandedPhases \, \wedge \, \forall \ Y,Z \in X. \ \relatedIntPwph^+(Y, Z) \big{\rbrace},
$$
where $\relatedIntPwph = \{ (X,Y) \mid \exists s \in X, t \in Y. \ \relatedIntPw(s, t) \}$ and $\relatedIntPwph^+$ is the transitive closure of $\relatedIntPwph$.
Informally, if processes can move via internal or \pairwise transitions between two expanded phases, then the two phases are merged to ensure that the processes always co-exist in the same phase.
\end{enumerate}
\end{definition}



This definition ensures that all the processes in the system are in the same phase at any given time in a program execution.\ourskip

\end{newtextt}




\para{\Chwellbehavedness Conditions.} 
Our {\em \chwellbehavedness} conditions ensure that all processes in a \aml system move in \phases such that the set of available events within a phase as well as the system's ability to move between phases (through globally-synchronizing events) is {\em independent of the number of processes}. Such independence is critical for decidability of MPVP, which needs to reason about an arbitrary number of processes.
In particular, since processes can only move between phases using globally-synchronizing events, these conditions ensure that such events behave in a way that is independent of the number of processes.
A \machine that satisfies these conditions is called \emph{\chwellbehaved}. 
In what follows,
we present the
\chwellbehavedness conditions.
\ourskip
We first define a classification of local transitions corresponding to globally-synchronizing events into {\em acting} and {\em reacting} transitions. For broadcasts, sending transitions are acting while receiving transitions are reacting. 
For the \PartitionCons primitive,
  winning transitions are acting while losing transitions are reacting.
For the \ValueCons primitive, transitions with winning proposals are acting while other transitions are reacting.
For each globally-synchronizing event $\act$, let $\trans{s}{A(\act)}{s'}$ (resp. $\trans{s}{R(\act)}{s'}$) denote a local acting (resp. reacting) transition of $\act$.
Additionally, for some event $\act$ and some subset $X$ of the local state space  $\LS$, we say that $\act$ is {\em \firable} in $X$ if some state in $X$ has an acting transition of $\act$.

\begin{definition}[\CHwellbehavedness Conditions]
\begin{compactenum}[(1)]
	\item[] 

	\item Every state $s \in P$ which has an acting transition $\trans{s}{A(\act)}{s’}$  must also have a 
	 corresponding reacting transition $\trans{s}{R(\act)}{s''}$.

    \item For each internal transition $\trans{s}{}{s’}$ that is accompanied by a
    reacting transition $\trans{s'}{R(\actb)}{s''}$ and for each state $t$ in the same phase as $s$, if event $\actb$ is \firable in that phase, then $t$ must have a path to a state with a reacting transition of event $\actb$.

  \item For each acting transition $\trans{s}{A(\act)}{s’}$ that is accompanied by a
  	reacting transition $\trans{s'}{R(\actb)}{s''}$ such that $\actb$ is \firable in the set 
  	\newww{
  	 $dst_\act$
  	   	 of \em destination states
  	  of event $\act$,}  
 (i) if there are other acting transitions $\trans{t}{A(\act)}{t’}$ for event $\act$, all of them must transition to a state $t'$ with a
 reacting transition $\trans{t'}{R(\actb)}{t''}$ of event $\actb$ and 
(ii) for every reacting transition $ \trans{u}{R(\act)}{u’}$ of $\act$, there must be a path from $u'$ to a state with a
reacting transition of event $\actb$.



\end{compactenum}
\label{def:wbcpaper}
\end{definition}

%
%
%
%
%

Intuitively, the conditions ensure that if a $\aml$ system with a given number of processes can move in phases, then any additional processes can go along with the existing ones by always taking a corresponding reacting transition. In particular,
condition (1) ensures that processes in the same state as the process taking the acting transition on some event $\act$ have a way to react to $\act$.
Condition (2) ensures that, if a process can reach a state with a reacting
transition on an \firable event in that phase, all other processes can also reach a state where they can 
react to that event.
Condition (3) ensures that
once a process takes an acting transition of event $\act$ and moves to a state $s'$ where a reacting transition of event $\actb$ can be taken, 
if $\actb$ is \firable in the phase of $s'$, 
all processes move in a way that ensures they can take a reacting transition of $\actb$ as well. \ourskip

\label{premissibleSpcsDef}
\para{\Permissible Safety Specification.} 
We target safety specifications which forbid the reachability of any global state where some number $m$ (or more) of processes are in some set of local states; simple instantiations of such specifications include mutual-exclusion and process-local safety properties.
Let $f$ be a Boolean formula over locations and \terma{int} variables of a \aml process. Let $f_i$ be $f$ indexed by the \pid $i$. For instance, $f = s.loc \neq c \ \wedge \ s.\sigma(v) < 1$ has the indexed formula $f_i = q[i].loc \neq c \ \wedge \ q[i].\sigma(v) < 1$. 
Let $ i_1, \ldots ,i_{m} $ represent distinct valuations over $\Ind$. Then, we define $\phi_{m,f}(n)$ as:
$$\spec{m}{f}{n} = \forall i_1, \ldots ,i_{m}. \; \neg \left( f_{i_1} \wedge \ldots \wedge f_{i_{m}} \right). $$
Intuitively, the formula $f$ encodes a set $\states(f) = \{ \ls \in \LQ \mid f = true\}$ of  process-local states (where $f$ holds)  and the property $\spec{m}{f}{n}$ forbids the reachability of a global state where $m$ or more processes are in the set of local states $\states(f)$. We call formulas of the form $\spec{m}{f}{n}$ \permissible safety specifications and note that all 
specifications in this paper can be expressed using this form. 
For example, the Distributed Store specification asserting that no more than 1 process is in location \code{Leader} is expressed as $\spec{2}{s.loc = \text{\code{Leader}}}{n}$, i.e., $\forall i_1, i_2. \; \neg \left(q[i_1].loc = \text{\code{Leader}} \wedge q[i_2].loc =  \text{\code{Leader}}\right)$. 

\neww{Examples of specifications that are \emph{not} \permissible include: ``there exists at least one process in location \code{Leader} at all times'', and, ``no more than half of the processes can be in location \code{Leader}''. The former forbids $m$ or less processes to be in a given set of local states (as opposed to $m$ or more) while the latter forbids the reachability of a possibly unbounded number of processes to a given set of states (as opposed to a fixed number $m$ of processes).
For the remainder of this section, we will focus on \permissible safety specifications of the form $\spec{m}{f}{n}$, but we note that our results extend to conjunctions and disjunctions of permissible safety specifications (see \appref{generalSpecs}).\ourskip}


%

\begin{newtextt}

On a high-level, if a \aml process is \chwellbehaved, then the behavior of the corresponding \aml system is independent of its number of processes. Hence, the reachability, or the lack thereof, of an error state corresponding to a violation of a \permissible safety specification is consistent across different ``sizes'' of the system. In other words, if an error state is reachable in a system with a given number of \chwellbehaved processes, adding additional processes will not render such an error state unreachable. 
Decidability follows from a similar argument in the opposite direction: if an error state is reachable in a system with some number of \chwellbehaved processes, then we can compute a number of processes sufficient for reaching the error state.


\end{newtextt}

The following theorem identifies the decidable fragment for \mpvp. 

\begin{theorem}
\label{thm:decidability}
\mpvp is decidable for \aml system process $P$ and \permissible safety specification $\phi(n)$ if:
\begin{enumerate}
  \item $P$ is \chwellbehaved.
  \item The state space of $P$ is fixed and finite\footnote{We note that this condition restricts the way participant sets of agreement primitives are built to the constant set \terma{All} or the result of a previous \PartitionCons instance $\consnamep$ (\termb{\consnamep}.\terma{winS} or \termb{\consnamep}.\terma{loseS}), hence ensuring the precondition of agreement is naturally met. In general, this condition can be relaxed to include some systems with an unbounded state space where such sets are built through communication.}. 
    \item There exists at most one \pairwise-receive transition per action per phase\footnote{Under full symmetry, this condition ensures that abstracting the receiver \pid (in any \pairwise-send transition $\sendrzid{s}{\actname}{\pid}{s'}$) does not introduce spurious behaviors.}.  
  
\end{enumerate}
\end{theorem}
\noindent{\em Proof Intuition.} The proof leverages a new fragment of an existing abstract model, the \hlm~\cite{GSP}, for which MPVP is decidable. 
The decidability result for the \hlm utilizes the framework of \emph{well-structured transition systems} (WSTSs). This entails defining a well-quasi-ordering (WQO) over the global state space of a GSP system as well as a set of sufficient ``well-behavedness'' conditions over the local GSP process definition to ensure that global transitions are ``compatible'' with the well-quasi-ordering. To admit a larger decidable fragment, we designed a novel WQO and relaxed these well-behavedness conditions. Further, we model process \failstop failures. We defer the intricacies of this extension (which we refer to as the \gspp), as well as the formal definitions of WSTSs, WQOs, and compatibility to \appref{gsppapp}. We note that, without this extension, the \chwellbehavedness conditions will not be initiability-aware (e.g., phase-compatibility condition (2) above will need to hold regardless of $\act$ being \firable or not).

We show that 
for any \aml process $\pl$ that satisfies the three conditions of \thmref{decidability}, one can construct a corresponding process $\php$ in \gspp such that there exists a simulation equivalence between their respective global state-transition systems and $\php$ belongs to the decidable fragment of \gspp. \neww{We refer the reader to \appref{decidabilityproof} for the full proof.}
\ourskip

Recall that, in \secref{related}, we discuss reasons why neither the  decidable fragment of  \hlm nor \gspp is directly suitable for designing agreement-based decidable systems.  \ourskip

\subsection{Cutoffs for Efficient Parameterized Verification}
\label{sec:cutoffs}

We define additional conditions on \aml programs to obtain small cutoffs and
enable efficient parameterized verification. These {\em \amenability conditions} ensure that
any global error state, where $m$ processes are in local states $\states(f)$ violating a permissible safety specification $\spec{m}{f}{n}$,
can be {\em reached} in a system with exactly $m$ processes iff it can be reached in a system of any size larger than $m$.
Thus, programs satisfying these conditions 
enjoy a small model property: $\spec{m}{f}{n}$ is satisfied in $\M(n)$ for all $n \in \Nat$ if $\spec{m}{f}{m}$ is satisfied in a system $\M(m)$ with a fixed number of processes $m$.
This requires the conditions to ensure that
 the reachability of a global state violating $\spec{m}{f}{m}$ in $\M(m)$ {\em does not depend on the existence of more than $m$ processes}. \ourskip 

\para{\AMenability Conditions.} 
We first define a notion of {\em independence} of transitions
and paths of a process. Informally, \freech transitions do not require
the \emph{existence} of other processes in certain states. For instance, in \PartitionCons \consagree, the winning transition $\ls \xrightarrow{\pcw(\chset,\chcard)}$ $\ls’$ is \freech since a winning process does not require the existence of a losing one to take that transition, but the losing transition $\ls$ $\xrightarrow{\pcl(\chset,\chcard)}$ $\ls’$ is not \freech since the losing process requires the existence of a winning process to take that transition. Note that acting transitions of globally-synchronizing events as well as internal transitions are \freech while reacting transitions are not \freech.
A path is independent if it consists of independent transitions.

\begin{definition}[\AMenability Conditions]
Let $P$ be a \chwellbehaved process, $\spec{m}{f}{n}$ a \permissible specification, and $\mathcal{F}$ the set of \freech simple paths from $\ls_0$ to a state $\ls \in \states(f)$
. We require either of the following to hold.\ourskip

\begin{compactenum}[(1)]
\item All paths from $\ls_0$ to $\states(f)$ are \freech, or, 
\item \neww{For every transition $\trans{\ls_s}{}{\ls_d}$ such that  $\ls_s \neq \ls_d$ and $\ls_s$ is a state in some path $p \in \mathcal{F}$, either 
    \begin{compactenum}[(a)]
    \item the state $\ls_d$ is in $p$ and the transition $\trans{\ls_s}{}{\ls_d}$ is \freech, or,
    \item the state $\ls_d$ is not in $p$ and all paths out of $\ls_d$ lead back to $\ls_s$ via \freech transitions.
  \end{compactenum}
}

\end{compactenum}
\label{def:amenabilitypaper}
\end{definition}


The conditions ensure that the processes required to {\em enable} a path to an error state
are available in $\M(m)$.
Condition (1) ensures that, if $m$ processes were to reach the error states, they can do so without requiring additional processes, since all paths to the error states are independent.
Condition (2) allows for some processes to ``diverge'' from the independent paths as long as they {\em return} independently. We note that the \kinarach tool  implements a more advanced version of this lemma that allows for more systems to have cutoffs.

We refer to the pair $\tuple{P,\spec{m}{f}{n}}$ as \emph{amenable} if 
$P$ is a \chwellbehaved \machine that satisfies the cutoff-amenability conditions w.r.t. \permissible safety specification $\spec{m}{f}{n}$.

\begin{newtextt}
On a high-level, an amenable pair $\tuple{P,\spec{m}{f}{n}}$ identifies systems where the minimum number of processes to trigger an error (i.e., $m$ process existing simultaneously in $\states(f)$) is, in fact, exactly $m$. This is achieved by ensuring that any path a process may take to an error state is \freech and hence if a process may reach an error state, it can do so without the help of other processes.
 
\end{newtextt}

\begin{lemma}
For an amenable pair $\tuple{P,\spec{m}{f}{n}}$, $c=m$ is a 
cutoff for \mpvp.
\label{lem:cutoffs}
\end{lemma}

\noindent{\em Proof Intuition}.
\neww{We utilize the cutoff results of \gspp. Using the construction in the proof of \thmref{decidability} to obtain a process $\php$ in the \gspp from a process $\pl$ in \aml, we show that if \amenability holds for $\pl$, then $\php$ will be \amenable 
and the resulting cutoff for $\tuple{\php,\spec{m}{f}{n}}$ is also a cutoff for $\tuple{\pl,\spec{m}{f}{n}}$. 
We refer the reader to \appref{cutoffapp} for the full proof.}
\ourskip

%
%
%
%
%

\para{Automation and Feedback.} 
While the phase-compatibility and cutoff-amenability conditions are somewhat intricate, 
we emphasize that our \kinarach tool automatically checks these conditions and additionally gives the system designer feedback on how to make a process \chwellbehaved 
and amenable.
This allows the designer to proceed without being caught up in the details of the exact conditions. 
\begin{newtextt}
The feedback varies depending on the failed condition and mainly aims to capture the root cause of the failure and to provide heuristically-ranked suggestions to fix it. 

A failure of a \chwellbehavedness condition can be succinctly captured by a phase, a set of local states in that phase, and set of acting/reacting transitions from these states over one or two events. This localization is valuable for the user to pin-point what changes are needed to render the system \chwellbehaved. \kinarach suggests edits that would eliminate the current violation and the user gets to pick which edit to implement. 

A failure of a \amenability condition for correctness properties $\spec{m}{f}{n}$ can be succinctly captured by a non-\freech path from the initial state $\ls_0$ to a state $\ls \in states(f)$. This path indicates a scenario where an error state could be unreachable in a system with $m$ processes but can be reachable in a bigger system; hence $m$ is not a valid cutoff. In these cases, \kinarach presents the non-\freech paths but does not suggest edits, and the user is responsible for eliminating the non-\freech transitions from these paths.
\ourskip
\end{newtextt}
\begin{newtext}

\noindent{\em Example.} 
Consider a system where a set of processes wish to select up to two processes that can then perform an action one after the other. The designer starts with the process definition shown in \figref{toy}. All processes start in location \termb{Start}, and coordinate to pick up to two processes to move to the \termb{Selected} location while the rest move to \termb{Idle}. From \termb{Selected}, the chosen processes send the \termb{getReady} broadcast and move to \termb{Prepare}. In \termb{Prepare}, they attempt to move to the \termb{Target} location one after the other using a \termb{sequencer} message. The correctness property for this system is $\spec{2}{loc = \code{Target}}{n}$ indicating that at most one process can be at the \termb{Target} location at any time.

\begin{figure*}[h]
\centering
\begin{tcolorbox}[colback=white,sharp corners,boxrule=0.3mm,top=1mm,bottom=1mm]
\hspace{1em}\begin{minipage}{0.47\textwidth}
\begin{ssdsl}
process SelectiveSerializer 
actions 
  br getReady : unit
  br sequencer : unit

 initial location Start (*\label{line:start}*)
   on (*\PartitionConsDSL{select}{All}{2}*)  (*\label{line:select}*)
     win: goto Selected
     lose: goto Idle

location Idle
   passive getReady, sequencer
\end{ssdsl}
\end{minipage}
\begin{minipage}{0.70\textwidth}
 \lstset{firstnumber=13}
\begin{ssdsl}
 location Selected
   on _ do
     sendbr(getReady)
     goto Prepare

location Prepare	
   on recv(sequencer) do 
     goto Target

location Target
  // perform action

\end{ssdsl}
\end{minipage}
\end{tcolorbox}

\caption{An Initial \aml Process Definition for Distributed Coordination for Serializing Access.}
\label{fig:toy} 
\end{figure*}

When \kinarach in run on this process definition, it reports that the system is not \chwellbehaved with the following feedback suggesting adding a receive handler of event \termb{getReady} from the \termb{Selected} location:\ourskip

\begin{tcolorbox}[colback=white,sharp corners,boxrule=0.3mm,right=0mm]
\footnotesize
\code{(Selected,\{\}) needs a corresponding reacting transition on getReady\\}
\code{Suggestions to solve this:}
\begin{compactenum}[ - ]
	\item \code{add transition (Selected,\{\})	------R(getReady)------>	(Prepare,\{\})}
	\item \code{add transition (Selected,\{\})	------R(getReady)------>	(Anywhere!,\{\})}
\end{compactenum}
\end{tcolorbox}

The designer accepts the first heuristically-ranked suggestion and adds the following handler to the \termb{Selected} location:
$$\code{on recv(getReady) do goto Prepare}$$

With this edit, the system is now \chwellbehaved. However, the system is not \amenable. \kinarach returns the following feedback:\ourskip

\begin{tcolorbox}[colback=white,sharp corners,boxrule=0.3mm,right=0mm]
\footnotesize
\code{Cutoff computation failed: on path}\\
\code{(Start,\{\})------A(select)------>(Selected,\{\})------A(getReady)------>(Prepare,\{\})------R(sequencer)------>(Target,\{\})}\\
\code{the following transition(s) are not \freech:}\\
\code{(Prepare,\{\})------R(sequencer)------> (Target,\{\})}
\end{tcolorbox}

Based on this feedback, the designer realizes that processes in \termb{Prepare} need to send the \termb{sequencer} broadcast to move to the \termb{Target} location, which is an \freech transition. Learning from the previous \chwellbehavedness violation, the user additionally adds the corresponding reacting transition. Thus, the designer replaces the receive handler from location \termb{Prepare} with the following two handlers: 
$$\code{on _ \ do sendbr(sequencer) goto Target}$$
$$\code{on recv(sequencer) do goto Prepare}$$
 
The system is now \chwellbehaved and \amenable, and \kinarach reports a cutoff value of two.\ourskip
\end{newtext}

\para{Modular Verification.} Recall that \lemref{abs-correctness} shows that any verified agreement protocol (i.e., one proven to satisfy agreement, validity, and termination) also meets the pre- and post-condition pair  of our agreement primitives. So, by verifying an \aml program with  agreement primitives w.r.t. to a safety specification, we can conclude that the program, when instantiated with any  agreement protocol that satisfies agreement, validity, and termination, also satisfies the safety specification. 
\ourskip

\section {Implementation and Evaluation}
\label{sec:evaluation}

We describe the implementation of 
\kinarach
\footnote{
A virtual machine containing \kinarach is publicly available \cite{zenodo}.}
	and evaluate the performance of its automated parameterized verification procedure on various benchmarks encoded in \aml.

\subsection{Implementation} 
\label{sec:impl}
\kinarach performs automated, parameterized verification of \aml programs in
three steps:
\begin{compactenum}[1.]
\item \noindent{\em Parsing.} 
\kinarach compiles \aml processes into the core fragment 
by rewriting all non-core handlers (e.g., handlers with if-statements or multiple send statements) into core handlers and expanding syntactic sugar.

\item \noindent{\em Analysis.} From the core fragment, \kinarach
  creates a labeled graph representing the process-level
  semantics including transitions that model \failstop failures.
\kinarach checks the \chwellbehavedness and \amenability 
conditions against this graph, and if the conditions are met, 
  computes a cutoff to verify the system.

\item \noindent{\em Verification.} \kinarach's verification engine is built on top of Kinara~\cite{Alur.AutomaticCompletionDistributed.X.2015}, a verification tool for distributed systems with a {\em fixed} number of processes. \kinarach extends Kinara to support the \PartitionCons and \ValueCons 
primitives as well as their global behaviors. 
\kinarach translates the core fragment of \aml into the input
representation accepted by the extended version of Kinara using the cutoff
number of processes, as computed during the analysis step. 
\neww{\Permissible safety specifications $\spec{m}{f}{n}$ are encoded in \kinarach as \termb{\dslatmost(}$m - 1$\termb{,\{loc:}\nont{bExp}\termb{\})} where the Boolean expression $f$ is such that
$\forall s \in states(f): s.loc = \code{loc} \wedge s.\sigma(bExp) = \code{True}$. For example, the property $\spec{3}{loc=\code{Replica} \wedge stored = 1}{n}$ is encoded as \termb{\dslatmost(}$2$ \termb{,\{Replica: stored == 1\}).}}
The environment process is automatically generated to nondeterministically send/receive all environmental communication actions that the \aml process expects. The specifications as well as the environment process are translated to Kinara's representation similarly.
\kinarach reports successful parameterized verification iff Kinara reports successful verification for the system consisting of the cutoff number of processes.
\end{compactenum}


\noindent{\em User Feedback.} 
\kinarach helps the user obtain a \chwellbehaved and \amenable process by providing heuristically ranked suggestions to handle any violation of the \chwellbehavedness and \amenability conditions
during the analysis step.
For instance, the \chwellbehavedness conditions do not hold if a local state has an acting transition but not its corresponding reacting transition. In this case, \kinarach returns the violated condition and suggests adding a handler that corresponds to the reacting transition from that state.

\subsection{Evaluation}
\label{sec:evalsubsection} 
The research questions we tackle in this evaluation are:
\begin{description}
	\item[RQ1] Can interesting agreement-based systems be modeled concisely in \aml?
	\item[RQ2] 
	Can interesting agreement-based systems be modeled in the decidable 
	fragment of \aml with relative ease? 
	
	\item[RQ3] Can \kinarach perform automated parameterized verification of agreement-based systems in \aml in a reasonable amount of time?
	
	\item[RQ4] Do \kinarach's cutoffs enable efficient verification?
\end{description}

In what follows, we first present our \aml benchmarks and address RQ1 and RQ2. We then analyze the performance of \kinarach on these benchmarks and address RQ3 and RQ4. All experiments were performed on an Intel Xeon machine with E5-2690 CPU and 32GB of RAM. \neww{We report the mean run time for 10 runs as well as the 95\% confidence interval for each benchmark.}\ourskip

\para{\aml Benchmarks.}
Our benchmarks are briefly described below.\footnote{\aml code for benchmarks available at: \href{https://tinyurl.com/m3zx7jxs}{https://tinyurl.com/m3zx7jxs}} 
\begin{compactenum}[1.]
	\item \noindent{\em Distributed Store} is the illustrative example from \secref{illustrative}.
	
	\item \noindent{\em Consortium} is a distributed system where a set of actors wants to reach a decision based on information the actors gather individually. A subset of the actors is elected and trusted with making a decision that is then announced to the rest of the actors. This resembles scenarios where a trade-off between trust and performance is needed (e.g., a consortium blockchain~\cite{libraref,hyperledgerref}). \neww{The safety property for this system is (1) at most two actors are elected to decide on a value for all processes and (2) that all actors agree on the decided value.}
	
	\item \noindent{\em Two-Object Tracker} is a system for collaborative surveillance based on leader election and is inspired by an  example in \cite{twoObjectTracker}. Upon detecting an object, a leader is elected to be responsible for monitoring it along with its followers. The system can additionally fork another set of processes to monitor a second object simultaneously. \neww{The safety property for Two-Object Tracker is that there can be at most two leaders at a time, and when a second object is spotted, each of the leaders is tracking a distinct object.}
	
	\item \noindent{\em Distributed Robot Flocking} is a distributed system where processes follow a common leader as a flock and is 
	inspired by an example in \cite{flocking}. \neww{Processes can disperse into various locations where they can elect a leader.}
	 The leader then issues directions to the rest of the flock. This is especially useful in self-stabilizing systems. \neww{The safety property for this system is that the system stabilizes by ensuring there can be at most one leader at a time making direction decisions.}
	
	\item \noindent{\em Distributed Lock Service} is a distributed lock service similar to Chubby \cite{burrows2006chubby} 
	for coarse-grained synchronization with an elected leader handling clients requests. \neww{Clients can interface
	with this lock service as a file system where they send reads and writes to an elected leader, and have their requests replicated safely on different servers. The leader periodically times out, sends a step down signal to the rest of the servers,  and a new round of election is used to pick a new leader. The safety property for this system is that at most one server is elected as a single point-of-contact for the clients.}
	
	\item \noindent{\em Distributed Sensor Network} is a sensor network
	application that elects a subset of processes, who have sensed an environmental signal, to report
	to a centralized monitor. \neww{In this application, the set of sensors that have detected the environmental signal (and hence need to coordinate) is dynamically built before invoking the agreement protocol. The safety property here is that the environmental signal is reported by no more than two sensors.}
	
	\item \noindent{\em Sensor Network with Reset} is a variant of the Distributed Sensor Network benchmark that
	uses a ``reset'' signal to resume monitoring for the environmental signal, thereby requiring an unbounded number of rounds of \consagree. \neww{The safety property is as before.}
	
	\item \noindent{\em Small Aircraft Transportation System (SATS) Landing Protocol } is the landing
	protocol of SATS proposed by ~\citet{satsref}.
		\neww{The goal of SATS is to increase access to small airports
	without control towers by allowing aircraft to coordinate with each other to
	operate safely upon entering the airport airspace.  For the landing protocol,
	the aircraft coordinate to choose successive subsets of aircraft to progress
	to the next phase of landing, until just one aircraft is chosen to land at a
	time. The desired safety properties for the SATS landing protocol, provided by NASA,
		are as follows: (1) there are a total of at most four aircraft across the airport vicinity; 
		(2) there are a total of at most two aircraft across each left (right) holding zone of the airport; and 
		(3) there is at most one aircraft that can attempt a final approach (i.e., attempt landing) at a time.}
	
	\item \noindent{\em SATS Landing Protocol II} is a version of the SATS landing protocol where aircrafts communicate explicitly to build a participant set when nearing the final approach, and return to a specific  holding zone if they miss landing. \neww{The safety property is as before.}	 
	
	\item \noindent{\em Mobile Robotics Motion Planning} is a distributed system based on an existing benchmark~\cite{Desai.DRONAFrameworkSafe.X.2017} \neww{where a set of robots share a workspace with obstacles, and need to coordinate their movements. The robots coordinate to create a motion plan by successively choosing each robot to create a plan while taking into account the previous robots' plans. The targeted property for this system is collision avoidance; this is achieved by allowing the robots to create their motion plans consecutively one-at-a-time.}

	\item \noindent{\em Mobile Robotics with Reset} is a variant of the Mobile Robotics Motion Planning benchmark that allows all the robots to return to their initial state upon receiving a signal to do so. \neww{The safety property is as before.}
	
	\item \noindent{\em Distributed Register} is a data store \'a la Atomix's AtomicValue~\cite{atomix} which gives a consistent view of a stored value under concurrent updates. \neww{Updates that do not change the stored data in the register are ignored. The safety property for the Distributed Register system is that no two replicas that are about to serve user requests may have different values of the register; hence ensuring the clients have a consistent view of the data.}
	
\end{compactenum}
\ourskip
\noindent{\em RQ1}. Our benchmarks are models of distributed agreement-based systems commonly found in the literature and have all been  encoded in \aml with relative ease by the authors of this work. Further, as can be seen in Column 2 of \tabref{resultss}, the corresponding \aml process definitions are fairly compact, i.e., within 100 lines of code (LoC). Thus, our answer to  RQ1 is Yes. 
 \ourskip
 
 \noindent{\em RQ2}. We found two factors valuable in addressing RQ2. 
%

\noindent{\em Value of User Feedback.} It is not always easy for a system designer to ensure that their initial model of a \aml process is \chwellbehaved. For example, when modeling the Distributed Lock Service benchmark, we made assumptions about the behavior of the system, causing us to omit reacting transitions on some events and, consequently, our initial model was not \chwellbehaved. However, the feedback provided by \kinarach helped identify the missing transitions that needed to be added.\ourskip

\noindent{\em Value of \gspp.} Prior decidability results did not encompass all of the benchmarks we evaluate; in particular, those marked with * in \tabref{resultss} fall outside prior known fragments. \gspp's extension of decidability results, on the other hand, enables decidable parameterized verification for all of our benchmarks.

Thus, with the help of \kinarach's user feedback and the \gspp decidable fragment, our answer to RQ2 is Yes. 
\ourskip
%

\begin{table}[t]
	\squeezecaption
	\squeezecaption
	\squeezecaption
	\squeezecaption
	\centering
	\caption{\kinarach Performance.}
	
	\begin{tabular}{l c c c r}
		\hline 
		\toprule
		\textbf{Benchmark}&\textbf{LoC}&\textbf{Phases}&\textbf{Cutoff}&\textbf{Time(s)} \\ 			
		\midrule 
		Distributed Store					&	64	&	2	&	3	&	$45.079 \pm 0.621$ \\
		Consortium$^*$						&	58	&	9	&	3	&	$6.953 \pm 0.022$ \\
		Two-Object Tracker$^*$				&	69	&	9	&	3	&	$0.641 \pm 0.006$ \\
		Distributed Robot Flocking			&	78	&	\neww{7}	&	2	&	$0.105 \pm 0.002$ \\
		Distributed Lock Service			&	38	&	2	&	2	&	$0.059 \pm 0.002$ \\
		Distributed Sensor Network			&	55	&	3	&	3	&	$1.041 \pm 0.003$ \\
		Sensor Network with Reset			&	63	&	\neww{3}	&	3	&	$1.662 \pm 0.012$ \\
		SATS Landing Protocol$^*$			&	90	&	3	&	5	&	$638.393 \pm 0.872$ \\
		SATS Landing Protocol II$^*$ 		&	99	&	\neww{5}	&	5	&	$736.417 \pm 3.659$ \\
		Mobile Robotics Motion Planning 	&	71	&	5	&	2	&	$0.114 \pm 0.004$ \\
		Mobile Robotics with Reset$^*$		&	83	&	4	&	2	&	$0.166 \pm 0.003$ \\
		Distributed Register				&	32	&	1	&	2	&	$0.329 \pm 0.006$ \\
		\bottomrule & 
	\end{tabular} 
	\label{table:resultss}
	\squeezecaption
	\squeezecaption	
	\squeezecaption
	\squeezecaption
	\squeezecaption
	\squeezecaption
	\reallysqueezecaption
	
\end{table}

 \noindent{\em RQ3}.
In \tabref{resultss}, for each benchmark we provide the number of phases, the cutoff used for verification, and the mean run time of \kinarach \neww{with its 95\% confidence intervals}. 
Notice that the cutoffs computed by \kinarach for all benchmarks are small (under 6 processes). 
Overall, \kinarach performs efficient
parameterized verification for all benchmarks,
 taking less than 2 seconds to verify most benchmarks, and about 12
	minutes for the largest benchmark, SATS Landing Protocol II. Thus, our answer to RQ3 is also Yes. 
\ourskip

\noindent{\em RQ4}. 
To examine the contribution of cutoffs in enabling efficient verification, 
we performed experiments studying the effect of varying the number of processes on the run time of \kinarach. As expected, increasing the number of processes causes the run time to grow exponentially. For instance, the time to verify the Consortium benchmark jumps from 9 seconds to about 8 minutes when verifying a system with 5 processes instead of the 3-process cutoff. Fortunately, \kinarach is able to detect small cutoffs to sidestep the exponential growth caused by increasing the number of processes, enabling practical parameterized verification. Thus, our answer to RQ4 is Yes.
%
%
\ourskip

\noindent{\em Remark.}
\kinarach additionally
	reports the number of phases, which correspond to global guards in
	the \gspp, that \kinarach automatically generates. This shows the value of
	automation as designing such guards manually is tedious and error-prone.
\ourskip

\section{Concluding Remarks}



We presented a framework, \kinarach, for modeling and efficient,
automated parameterized verification of agreement-based 
systems.  The framework supports a modular approach to
the design and verification of distributed systems in which systems are (i) modeled using sound abstractions of complex distributed components and (ii) verified using model checking-based techniques assuming that the complex components are verified separately, presumably using deductive techniques.

In ongoing work, we focus on extending \kinarach to handle non-blocking
communication and network failures using ``channels'' that can buffer or drop messages, 
to support infinite variable domains using abstract interpretation, 
and to help system designers {\em synthesize} 
amenable processes.

Eventually, we hope to see this framework generalized by us or our readers to 
other verified distributed components and richer properties such as liveness. 
We also hope to see more conversations and verification frameworks, in particular layered ones, 
that cut across the deductive verification and model checking communities.


\section*{Acknowledgments}
We thank Ilya Sergey, Rupak Majumdar, Isil Dillig, Thomas Wahl, and Marijana Lazic for their invaluable feedback on various drafts of this paper, and Derek Dreyer for helping us interpret reviewer comments about an earlier draft. We are grateful to Shaz Qadeer and Ken McMillan for their thought-provoking questions at earlier stages of this work that helped shape this paper. We also thank Abhishek Udupa for patiently answering all our questions about the Kinara tool. Finally, the authors are grateful to the anonymous reviewers from POPL 2019, CAV 2019, POPL 2020, CAV 2020, POPL 2021, PLDI 2021, and OOPSLA 2021 who chose to take the time to provide constructive feedback on our submissions. This research was partially supported by the National Science Foundation under Grant Nos. 1846327, 1908504, and by grants from the Purdue Research Foundation and Amazon Science. Any opinions, findings, and conclusions in this paper are those of the authors only and do not necessarily reflect the views of our sponsors.

\newpage
%
\bibliographystyle{ACM-Reference-Format}
\bibliography{ms}

\newpage

\appendix
\section{Symmetry For Parameterized Correctness}\label{app:symmetry}

Since {parameterized} systems are designed to work for an arbitrary number of processes, their behaviors should be independent of a specific \pid (as such \pid may not even exist in every instantiation of the system). As a result, such parameterized systems naturally exhibit many similar global behaviors. In this section, we define the notion of full symmetry, how to check if a system is fully-symmetric , and the effect of full symmetry on verification. 

\subsection{Symmetry Reduction}
\label{app:basicsymmetry}

\para{Full Symmetry.} 
Let $\pi: \{1,\ldots,n\} \rightarrow
\{1,\ldots,n\}$ be a permutation acting on the set $\Ind$ of process indices. 
Let $\PG$ denote the set of all permutations over $\Ind$.
A permutation of a global state 
$\gq = (s_0, s_1, \dots, s_n)$ can then be defined as: $\pi(\gq) =
(\pi(\loq_0), \pi(\loq_1), \dots, \pi(\loq_n))$, where $\pi(\loq_i) =
(\q_{\pi(i)},\pi(\sigma_i))$ for $\loq_i=(\q_i,\sigma_i)$. Note that $\pi(\sigma_i)$ depends on the type of
the local variable being permuted: if it is of type $\Ind$, then $\pi(\sigma_i) = \sigma_{\pi(i)}$, otherwise
$\pi(\sigma_i) = \sigma_i$~\footnote{In case the variable was of an enumerated type (e.g., set, array, or record) containing values of type $\Ind$, then the permutation is applied recursively to all elements. If the array has an index type $\Ind$, then we permute the array elements
  themselves, too.}.

\begin{definition}[\cite{emerson2003combining,wahl2007adaptive}] \label{def:fullsym}
A global transition system $\M$ composed of identical processes 
with index set $\Ind$ is \emph{fully-symmetric} 
if its transition relation $\GT$ is invariant under permutations in $\PG$:
$\forall \pi \in
\PG: \, \pi(\GT) = \GT$, where
$\pi(\GT) = \{(\pi(\gq_1), \pi(\gq_2)): (\gq_1,\gq_2) \in \GT\}$.
\end{definition}

\aml syntax enforces sufficient syntactic constraints on manipulation of \pids to ensure full symmetry. Such constraints are similar to these in~\cite{emerson2003combining,wahl2007adaptive,emerson2005dynamic} where they prove that limiting predicates over \pids to equality and diseqality checks yields full-symmetry of $\M$.\\
\para{Verification Advantages of Full Symmetry.} 
Emerson and Sistla~\cite{emerson1996symmetry} show that it is possible 
to exploit the symmetries present in a global transition system $\M$ (of a system with
many similar processes) to improve scalability of model checking by constructing a compressed 
{\em quotient} structure $\aqs$ such that 
$\M \models \phi \iff \aqs \models \phi$, where $\phi$ is 
any ($\text{CTL}^*$) specification. 
It follows from their result that $\aqs$ 
can be constructed for any $\M$ that is fully-symmetric 
and can enable symmetry reduction for model checking 
w.r.t. any LTL specification.
We refer the
interested reader to Emerson and Sistla \cite{emerson1996symmetry} for further details.

\subsection{Symmetry of \aml Programs} \label{app:chsymmetry}
Since we introduced the \consagree primitives \PartitionCons and \ValueCons, we need to show that full symmetry is preserved. In a global transition
system, $\M_{\ch} = (\GQ,q_0,\GT)$ of a \aml program, let $\GT_{\ch}$ denote
the set of all \consagree transitions in $\GT$.  
Let
$\M = (\GQ,q_0,\GT\setminus \GT_{\ch})$ denote the global transition system
without the agreement transitions of $\M_{\ch}$.
\begin{lemma}
\label{lem:indexIndepOfChApp}
If $\M$ is fully-symmetric, then $\M_{\ch}$ is fully-symmetric.
\end{lemma}
\begin{proof}

Essentially, we need to show that $\GT_{\ch}$ is also invariant under permutations in $\PG$, i.e., $$\forall \pi \in \PG, (\gqs, \gqe^\winset) \in \GT_{\ch}: 
\, (\gqs, \gqe^\winset) \in \GT_{\ch} \; \iff \; (\pi(\gqs),\pi(\gqe^\winset)) \in \GT_{\ch}$$

We first examine \ValueCons transitions. Recall that a \ValueCons transition is created between $\gqs$ (that encodes the set of participants $\chset$, the desired number of values to agree on $\chcard$, and proposal variable $\chvar$) and a possible $\gqe^\winset$ (that encodes a winning set of values $\winset \in \winset^*$).
We prove this by contradiction. Assume: 
	$$\exists \pi \in \PG:
	\, (\gqs, \gqe^\winset) \in \GT_{\ch} \; \wedge \; (\pi(\gqs),\pi(\gqe^\winset)) \not \in \GT_{\ch}$$

\neww{Let $\gqs' = \pi(\gqs)$. Note that we obtain a new participant set $\chset' = \pi(\chset)$. 
Since constructing $\winset^*$ is invariant to permutations, we have  ${\winset^*}' = \pi(\winset^*)$. Because $\gqs'$ is also reachable for any $\pi$ ($\M$ is fully symmetric), we have a \ValueCons transition $(\gqs',\gqe^{\winset'}) \in \GT_{\ch}$ out of that state for every $\winset' \in {\winset^*}'$. Hence, we have $(\gqs',\gqe^{\winset'}) \in \GT_{\ch}$ for $\winset' = \pi(\winset)$ where $\gqe^{\winset'} = \pi(\gqe^{\winset})$. So, we have a contradiction, and:}
$$\forall \pi \in \PG:  (\gqs,\gqe^\winset) \in \GT_{\ch} \iff (\pi(\gqs), \pi(\gqe^\winset)) \in \GT_{\ch}$$
For \PartitionCons transitions, the essence of the argument is identical. 
Since both $\GT_{\ch}$is invariant under permutations in $\PG$, we conclude that $\M_{\ch}$ is fully-symmetric.
\end{proof}

\section{The \gspp Model}
\label{app:gsppapp}
In this section, we present, \gspp, our extension to the \hlm~\cite{GSP}. Both models share the same representations of local and global semantics, but use different conditions to classify systems for which the parameterized verification problem is decidable. We first present the \hlm in \appref{gspapp} and \gspp in \appref{gsppsec}.

\subsection{The \gbc Model \cite{GSP}}
\label{app:gspapp} 

The \hlm generalizes synchronization-based models including models based on rendezvous and broadcasts.  In the \hlm, each global transition synchronizes
all processes, where multiple processes {\em act} as the
senders of the transition, while the remaining processes react as
receivers. The model supports two types of transitions: (i) a
\emph{$k$-sender transition}, which {\em can fire} 
only if the number of
processes available to act as senders 
in the transition is at least $k$ and {\em is fired} with exactly $k$ processes 
acting as senders, and, (ii)
a \emph{$k$-maximal transition}, 
which {\em can fire} only if the number $m$ of
processes available to act as senders is at least one 
and {\em is fired} with $k$ processes acting as senders if $m \geq k$, or, with 
$m$ processes acting as senders, otherwise. 
Additionally, each
transition can be equipped with a {\em global guard} which identifies a subset of the
local state space. A transition is {\em enabled} whenever 
it can fire and 
the local states of all processes 
are in the transition guard.\ourskip

\para{Processes.} A \gbc process is defined as
$\ph=\tuple{A,\LS,\ls_0,T}$, where $A$ is a set of \emph{local} actions, $\LS$
is a finite set of states, $\ls_0 \in \LS$ is the initial state, and $T
\subseteq \LS \times A \times \LS$ is the transition relation. The set $A$ of
local actions corresponds to a set $\GActions$ of \emph{global} actions.  Each
global action $a \in \GActions$ has an arity $k$ with local send actions
$a_1!!,\ldots,a_k!!$ and a local receive action $a??$, and 
is further associated with either a $k$-sender or a 
$k$-maximal global transition
as well as a global guard $\guard_a \subseteq \LS$ of the transition. \ourskip

\para{Composition of Processes.} Given a process
$\ph$, the parameterized global transition system is
defined as $\Mh(n)=\tuple{Q,\conf_0,R}$ where 
$Q = \{0,\ldots,n\}^{|\LS|}$ is the set of states, 
$\conf_0$ is the initial state and 
$R$ is the global transition relation. Thus, 
a global state $\conf \in Q$ 
is a vector of natural numbers, representing the number of
processes that are in any given local state $\ls \in \LS$.
The global transition relation $R \subseteq Q \times \GActions \times 2^\LS
\times Q$ defines how processes synchronize using a $k$-sender or a $k$-maximal
transition to move between global states. In a global transition based on a
global action $a \in \GActions$ with arity $k$, each of the local send actions
$a_1!!,\ldots,a_k!!$ is taken by \emph{at most} one process. A transition $\trans{\conf}{a,G_a}{\conf'}$ based on a $k$-sender 
action is in $R$ if (i) the transition is enabled (i.e. all the processes in $\conf$ are in the subset of the local state space 
defined by the transition guard $G_a$), and (ii) in $\conf$, there are at least $k$ senders each taking a local sending 
transition $\trans{\ls}{a_i!!}{\ls'}$. The remaining processes take the local transition $\trans{\ls}{a??}{\ls'}$. The global 
state $\conf'$ is obtained by all processes moving accordingly. Transitions $\trans{\conf}{a,G_a}{\conf'}$ based on a $k$-maximal 
action behave similarly, except that at least one (instead of $k$) sender is required, and $\conf'$ is obtained by maximizing (up 
to $k$) the number of processes that can act as senders. For each global action $a$, the \hlm defines a synchronization matrix $M_a$ that describes how receivers move and a (set of) 
vector-pairs $(\textbf{v}_a,\textbf{v}_a')$ that describe how senders move. In particular, $M_a(\ls,\ls')=1$ if there is a local 
transition $\trans{\ls}{a??}{\ls'}$,
  $\textbf{v}_a$ dictates the expected senders each state $t \in \LS$: $\textbf{v}_a(t)= | \{
  \trans{\ls}{a_i{!!}}{\ls'} \mid \ls = t \}|$, and the vector
  $\textbf{v}_a'$ describes the number of senders that will be in each state $t \in \LS$ after the
  transition: $\textbf{v}_a'(t)= | \{ \trans{\ls}{a_i{!!}}{\ls'} \mid \ls' = t \}|$. Then, a global state $\conf'$ resulting from 
  transition $\trans{\conf}{a}{\conf'}$ can be computed as: $\conf' = M_a \cdot (\conf - \textbf{v}_a) + \textbf{v}_a'$.\ourskip

\para{Well-Structured Transition Systems (WSTSs).} \citet{Finkel87} introduces the WSTSs framework used to prove decidability for 
reachability problems. A WSTS $M(n)=\tuple{Q,R,\wqo}$ is a transition system with a set of infinite states $Q$, a transition 
relation $R$ over $Q$, and a well-quasi-ordering (WQO) $\wqo$ over $Q$ that is compatible with the transition relation $R$. 
Compatibility is defined as follows: for every $\conf, \conf', \altconf \in Q$ with $\conf \wqo
\altconf$ and $\conf \rightarrow \conf'$ there exists $\altconf' \in Q$ with
$\conf' \wqo \altconf'$ and $\altconf \rightarrow^* \altconf'$.\ourskip

\para{Upwards-Closed Sets.} A possibly infinite subset $U \subseteq Q$ is called upwards-closed w.r.t. $\wqo$ if $\conf \in U$, 
then every $\altconf$ such that $\conf \wqo \altconf$, $\altconf \in U$. Every upwards-closed set $U$ has a finite set of minimal 
elements w.r.t. $\wqo$ which are called the basis of $U$.\ourskip

\para{Effective Computability of Predecessors.} For $U \subseteq Q$, let $Pred(U)$ denote the 
predecessor states of $U$ with respect to $R$. We say $Pred(U)$ is effectively computable if there exists an algorithm that 
computes a finite basis of $Pred(U)$ from any finite basis of any upwards-closed $U$.

The following theorem states the decidability result of WSTSs.
\begin{theorem}[\cite{FS01}]
In a WSTS with effectively computable $Pred$, reachability of any upwards-closed set is decidable.
\end{theorem}

\para{Parameterized Verification in the \gbc Model.} \citet{GSP} define
a set of {\em well-behavedness} conditions over the process definition $\ph$ as well as a WQO over the global states to
ensure decidability of parameterized verification; the decidability result
itself is based on a reduction to WSTSs\cite{Finkel87}. 
\citet{GSP} uses the following WQO: 
$$\conf \altwqo \altconf \; \text{ iff } \; \left( \conf \wqo \altconf \land \forall \guard \in \mathcal{G}: \left( \supp(\conf) \subseteq \guard \iff \supp(\altconf) \subseteq \guard \right) \right),$$
where
\begin{compactenum}

\item $\wqo$ is the component-wise order on global state vectors 
$\conf$, $\altconf$:
$\conf \wqo \altconf \; \text{ iff } \; \conf(\ls) \leq \altconf(\ls) \text{ for all } \ls \in \LS,$
\item $\mathcal{G}$ is the set of all guards in the system, and,
\item $\supp(\conf) = \{ \ls \in \LS 
\mid \conf(\ls) > 0 \}$ is the support of a global state $\conf$, i.e., the set of local states that appear at least once in $\conf$.
\end{compactenum}

Intuitively, $\altconf$ is 
greater than $\conf$ if $\altconf$ has at least as many 
processes as $\conf$ in any given state, and for every transition 
$\trans{\conf}{a}{\conf'}$ that is enabled in 
$\conf$, a transition on action $a$ is also enabled in $\altconf$. 

An example of a well-behavedness condition that \citet{GSP} impose on the process definition is the following.
For a $k$-sender action $a$ with local sending transitions $\trans{\ls_i}{a_i!!, \guard}{\ls_i'}$ for $i \in \{1,\ldots,k\}$, let $\senderset_a$ be the set of all states $\ls_i$, $\senderset_a'$ the set of states $s_i'$, and $M_a$ the synchronization matrix. 
We say that action $a$ is strongly guard-compatible if the following holds for all $\guard' \in \guardset{:}$
\[ 
\senderset_a' \subseteq \guard' \Rightarrow \forall \ls \in \guard{:}~ M_a(\ls) \in \guard' 
\tag{C1} \label{eq:guard-comp-k-sender}
\]

and weakly guard-compatible if the following condition holds: 

\[ 
\senderset_a' \subseteq \guard' \Rightarrow \forall \ls \in \guard{:}~ \left( M_a(\ls) \in \guard' \lor \exists \ls' \in \LS: (\ls' \triangleleft \senderset_a' \land M_a(\ls) \rightsquigarrow \ls') \right).
\tag{C1w} \label{eq:guard-comp-k-sender-weak}
\]

\citet{GSP} present additional conditions for phase-compatibility of $k$-maximal transitions, which we omit here for brevity.

\subsection{\gspp}
\label{app:gsppsec}

In this section we present a novel \emph{firability-aware} WQO which we define over the global states. We start with some definitions. We define a relation $\text{assoc}$ to connect actions to their relevant guards: $\text{assoc}(a, G) \iff \exists s \xrightarrow{a!!, G} t$. 
Additionally, let $\mathcal{AG}$ be the set of all associated actions and guards: $\mathcal{AG} = \{(a, G) \in \mathcal{A} \times \mathcal{G}\mid \text{assoc}(a, G)\}$.
Moreover, we define the following function to capture when an action $a$ can fire from global state $\conf$ using the send-vector $\textbf{v}_a$: $\text{canFire}(\conf,\textbf{v}_a) \iff \forall s \in S. \conf[s] \geq \textbf{v}_a[s]$. For simplicity, we also lift this existentially to actions instead of their associated send-vectors: $\text{canFire}(\conf, a) \iff \exists \textbf{v}_a.\ \text{canFire}(\conf,\textbf{v}_a)$.

We define a function below to capture the set of actions which are enabled and can fire in a global state. We say that these action-guard pairs are "ready" in the global state $\conf$.
$$\text{ready}(\conf) = \{(a, G) \in \mathcal{AG}\mid 
\text{canFire}(\conf, a)\ \land\ 
\text{supp}(\conf) \subseteq G  
\}$$
Using this function, we can define the firability-aware WQO\footnote{We show that $\newwqo$ is a WQO by proving that every infinite sequence of global states $\conf_1, \conf_2, \ldots$ contains $\conf_i, \conf_j$ with $i<j$ and $\conf_i \newwqo \conf_j$. To this end, consider an arbitrary infinite sequence $\bar{\conf} = \conf_1, \conf_2, \ldots$. Then there is at least one set $\mathbb{A}$ of pairs $(a, G)$ such that infinitely many $\conf_i$ have $\text{ready}(\conf_i) = \mathbb{A}$ (since there are finitely many guards and finitely many actions). Let $\bar{\conf'}$ be the infinite subsequence of $\bar{\conf}$ where all elements have $\text{ready}(\conf_i') = \mathbb{A}$. Since $\preceq$ is a WQO, there exist $\conf_i', \conf'_j$ with $i < j$ and $\conf_i' \preceq \conf'_j$, and since $\text{ready}(\conf_i') = \text{ready}(\conf_j') = \mathbb{A}$, clearly $\text{ready}(\conf_i') \subseteq \text{ready}(\conf_j')$, so we also get $\conf_i' \newwqo \conf_j'$. Since $\conf_i' =\conf_k$ and $\conf_j' =\conf_l$ for some $k < l$, we get $\conf_k \newwqo \conf_l$ for $k < l$, and thus $\newwqo$ is a WQO.} as follows:
$$\conf \newwqo \altconf \iff (\conf \preceq \altconf\ \land\ \text{ready}(\conf) \subseteq \text{ready}(\altconf))$$
Intuitively, we say that $\conf \newwqo \altconf$, $\altconf$ has as many processes in every local state in $\conf$, and if every enabled firable action in $\conf$ is enabled and firable in $\altconf$.

With firability-awareness in mind, we relax condition (C1) as follows. Let $M_{a}(G) \coloneq \{M_{a}(s) \mid s \in G \}$, i.e., the set of all states where the receivers of $a$ that started in $G$ transition to. Then, condition (C1) becomes:
\[ 
\forall (a1,G1),(a2,G2) \in \mathcal{AG}: \left( (\senderset_{a1}' \subseteq G2) \wedge (((\senderset_{a1}' \cup M_{a1}(G1)) \cap \senderset_{a2}) \neq \varnothing) \right) \Rightarrow \forall \ls \in G1: M_{a1}(\ls) \in G2
\tag{C1r} 
\]

%

In a nutshell, the state of guard $G2$ after $a1$ happens is irrelevant if all the processes performing $a1$ move outside of the sending set of $a2$ as, at that point, $a2$ is not firable anyway. Similarly, Condition (C1w) can be relaxed as follows:
\begin{multline}
\forall (a1,G1),(a2,G2) \in \mathcal{AG}: 
\left( (\senderset_{a1}' \subseteq G2) \wedge (((\senderset_{a1}' \cup M_{a1}(G1)) \cap \senderset_{a2}) \neq \varnothing) \right) \Rightarrow \\
\left(\forall \ls \in G1: \left( M_{a1}(\ls) \in G2 \lor \exists \ls' \in \LS: (\ls' \triangleleft \senderset_a' \land M_{a1}(\ls) \rightsquigarrow \ls') \right)
\right)
\tag{C1wr} 
\end{multline} 

Other conditions in \cite{GSP} can be relaxed similarly.

\subsection{Parameterized Verification in the \gspp}
In this section, we show that, under the relaxed guard-compatibility conditions and the new WQO $\newwqo$, we obtain the following theorem. Let $M_\infty$ denote the global transition system composed of an infinite number of $\php$ processes.

%
%

\begin{theorem}
If $M_\infty$ is based on a well-behaved \gspp process $\php$, then $M_\infty$ is a WSTS and we can effectively Compute Pred.
\label{thm:deciability}
\end{theorem}

\begin{proof}
We show compatibility of transitions w.r.t. $\newwqo$, i.e., if $\conf \newwqo \altconf$ and $\conf \rightarrow \conf'$, the $\exists \altconf'$ with $\conf' \newwqo \altconf'$ and $\altconf \rightarrow^*\altconf'$. We consider the case of $k$-sender transitions here. $k$-maximal transitions are handled similarly.

Suppose $a1$ is a $k$-sender action. Let $\conf \xrightarrow{a1, G1} \conf'$ be a transition and $\conf \newwqo \altconf$. Since $\conf \xrightarrow{a1, G1} \conf'$, $(a1, G1) \in \text{ready}(\conf)$, since $\conf \newwqo \altconf$, this implies that $(a1, G1) \in \text{ready}(\altconf)$, we know that transition $\altconf \xrightarrow{a1, G1} \altconf''$ is possible and $\altconf''$ can reach $\altconf'$ with zero or more internal transitions (i.e.,$\altconf''\rightarrow^{*} \altconf'$), and by the proof of Theorem 2~\cite{GSP} we know that $\conf' \wqo \altconf'$. To prove compatibility with respect to $\newwqo$, it remains to show that $\forall (a2, G2) \in \mathcal{AG}: (a2, G2) \in \text{ready}(\conf') \implies (a2, G2) \in \text{ready}(\altconf')$.

First assume that condition (C1r) holds. Then, let $(a2, G2) \in \mathcal{AG}$ be an arbitrary associated action-guard pair. By this condition, we either have
\begin{compactenum}
\item $\senderset_{a1}' \not \subseteq G2$, i.e., at least one sender of $a1$ moves to a state outside of $G2$, so we are guaranteed that $G2$ is disabled after $a1$ fires. In this case, $\text{supp}(\conf') \not \subseteq G2$, so $(a2, G2) \not \in \text{ready}(\conf')$; therefore $(a2, G2) \in \text{ready}(\conf') \implies (a2, G2) \in \text{ready}(\altconf')$ is (trivially) satisfied.
\item $((\senderset_{a1}' \cup M_{a1}(G1))\cap \senderset_{a2} = \varnothing)$, i.e., all senders and receivers of $a1$ move into states with no sending transition on $a2$. In this case, $ \text{canFire}(\conf', a2)$ is false, so $(a2, G2) \not \in \text{ready}(\conf')$; therefore $(a2, G2) \in \text{ready}(\conf') \implies (a2, G2) \in \text{ready}(\altconf')$ is (trivially) satisfied.
\item $\senderset_{a1}' \subseteq G2 \land \forall \ls \in G1: M_{a1}(\ls) \in G2$, i.e., all potential senders move into $G2$, and all receivers move into $G2$ 
to arrive in a state where all processes are in $G2$. Additionally, if $(a2, G2) \in \text{ready}(\conf')$ it must be the case that $\text{canFire}(\conf', \textbf{v}_{a2})$ for some $\textbf{v}_{a2}$, and since we know that $\conf' \wqo \altconf'$ then
we have that $\text{canFire}(\altconf', \textbf{v}_{a2})$\footnote{
		We know that $\forall s \in S: \altconf'[s] \ge \conf'[s] \wedge \conf'[s] \geq \textbf{v}_{a2}[s]$ and hence, $\forall s \in S:\ \altconf'[s] \geq \textbf{v}_{a2}[s]$.}; therefore $(a2, G2) \in \text{ready}(\altconf')$.
\end{compactenum}
Hence, because $\conf' \wqo \altconf'$ and $\forall (a2, G2) \in \mathcal{AG}: (a2, G2) \in \text{ready}(\conf') \implies (a2, G2) \in \text{ready}(\altconf')$, we conclude that $\conf' \newwqo \altconf'$.

In case condition (C1rw) holds, the argument is the same, except that the receivers in case (3) can take multiple internal transitions to reach $\altconf'$.

%
%

Effective computability of $Pred$ follows from the proof of Theorem 2~\cite{GSP}. The only difference is that we must consider the guards, i.e., a predecessor is only valid if it additionally satisfied the guard of the transition under consideration.
\end{proof}

\neww{\para{Cutoff Results.} All the cutoff results from the \hlm naturally extend to \gspp. In essence, the additional special transitions that model \failstop failures are \freech as they never require additional processes to fire. Furthermore, the special crash state $\crashstate$ is never on a path to a state that shows in the specifications. Combined with the extended decidable fragment, this entails a larger cutoff-yielding fragment in \gspp.}

\section{Decidable Parameterized Verification for \aml Programs}
\label{app:decidabilityproof}
In this section, we 
prove that if a process is \chwellbehaved, then the \mpvp is decidable.

\begin{theorem}
\label{thm:decidabilityapp}
	\mpvp is decidable is decidable for \aml system process $P$, and \permissible safety specification $\phi(n)$ if:
	\begin{enumerate}
		\item $P$ is \chwellbehaved.
		\item The state space of $P$ is fixed and finite.
		\item There exists at most one \pairwise-receive transition per action per phase.
	\end{enumerate}
\end{theorem}

For the remainder of this section, we lay out the proof of \thmref{decidabilityapp}.
To show that parameterized verification is decidable for \chwellbehaved programs in \aml, we will utilize decidability and cutoff results of \gspp by showing that for each process $\pl$ that satisfies the above conditions, there exists a corresponding process $\php$ in the \gspp such that there exists a simulation equivalence between $\pl$ and $\php$.

\subsubsection{Simulation Equivalence}
We now present a mapping procedure (Rewrite) in \algoref{mapping} and show that there exists a simulation equivalence between a a rewritable \chwellbehaved process $\pl$ and the corresponding $\php = $Rewrite($\pl$). The rewriting procedure consists of a series of rewriting steps in which the semantics of each type of transition of a \aml program is converted into (a set of) transitions in \gspp.\ourskip

\para{\textsc{Rewritable}.} This function checks if $\pl$ has a fixed and finite local state space, and that the pairwise transitions are used in a way that does not violate full symmetry when the process IDs are abstracted away. Under these conditions, we show that any \chwellbehaved $\pl$ can be mapped to a \wellbehaved $\php$.

Recall that the set $\LQ$ represents the local state space of $\pl$. We start with a processes definition, $\php$, in \gspp whose local states are $\LQ$. In $\ml$, a global state $q_{\merc} \in \GQ_{\merc}$ of the form $\LQ^\n$ is a concatenation of the local states of all processes. In the $\gspp$, a global state $\conf_{Core} \in \GQ_{Core}$ of the form $\Nats^{|\LQ|}$ is a counter representation recording how many processes are in a given local state. 
Below, we provide an abstraction function that maps the global state space of $\ml$ to the global state space of $\mhp$. We define the function $\absfn: \GQ_{\merc} \rightarrow \GQ_{Core}$ 
as follows:
$\conf_{Core}(\loq) = \sum_{\loq_i \in \gq_{\merc}} \indicatorr{\loq_i = \loq}, $
 where $\indicatorr{cond}$ evaluates to 1 if $cond$ is \true~and 0 otherwise. The function counts the local states in $\gq_{\merc}$ and encodes that into the counter representation in $\conf_{Core}$.

\begin{algorithm}[t]
\footnotesize
  \SetKwFunction{mapproc}{Rewrite}
  \SetKwProg{myproc}{procedure}{}{}
  \myproc{\mapproc{$\pl$}}{
  \SetKwInOut{Input}{Input}
  \SetKwInOut{Output}{Output}
  \Input{$\pl$, a \chwellbehaved \aml process}
  \Output{$\php$, a process in \gspp}
  \BlankLine

 
\If {\Call{Rewritable}{$\pl$}} {
  $\php = $ \Call{RewriteBroadcastTransitions}{$\pl$}  \label{line:broadcast}\\
  $\php = $ \Call{RewriteRendAndInternalTransitions}{$\php$}  \label{line:rendezvous}\\
  $\php = $ \Call{RewritePartitionTransitions}{$\php$}  \label{line:wam}\\
  $\php = $ \Call{\RewriteValueConsTransitions}{$\php$}  \label{line:valstore}\\
  
  $\php = $ \Call{RewriteCrashTransitions}{$\php$}  \label{line:crash}\\
  \Return $\php$
}
}
\caption{Procedure for rewriting a process $\pl$ to a process $\php$.}
\label{algo:mapping}
\end{algorithm}
\para{\textsc{RewriteBroadcastTransitions}.}
Consider an arbitrary broadcast action $\actname$. For each broadcast send transition $\sendbr{\loq} {\actname}{\loq'}$ in $\pl$, we create a 1-sender transition $\trans{\loq}{\actname_1!!, src_\actname}{\loq'}$ in $\php$, and for each broadcast receive transition $\recvbr{\loq}{\actname}{\loq'}$ in $\pl$, we create a receiving  transition $\trans{\loq}{\actname??, src_\actname}{\loq'}$ in $\php$. Recall that $src_\actname$ is the source set of action $\actname$. We add the guard $src_\actname$ to ensure that the broadcast primitives have similar semantics. Hence, it is not hard to see that the following correspondence holds:

\begin{lemma}
\label{lem:broadcasts}
Let $\actname$ be a broadcast action, and let $G$ be a guard for $\actname$ obtained by the \textsc{RewriteBroadcastTransitions} procedure described above. Then,
$
\forall q_1,q_2 \in \GQ_{\merc}: \trans{q_1}{\actname}{q_2} \in \ml \iff 
\trans{\absfn(q_1)}{\actname,G}{\absfn(q_2)} \in \mhp.
$

\end{lemma}
\para{\textsc{RewriteRendAndInternalTransitions}}
Internal and \pairwise transitions can be supported using 1- and 2-sender actions, respectively. We replace each internal transition $\trans{\loq}{\epsilon}{\loq'}$ in $\pl$ with a 1-sender transition $\trans{\loq}{\phase(\loq)}{\loq'}$ in $\php$, where $\phase(s)$ is the union of all phases that contain the state $s$. We also replace \pairwise transitions $\sendrz{\loq}{\actname}{\loq'}$ and  $\recvrz{\loq} {\actname}{\loq'}$ with 2-sender transitions $\trans{\loq}{\actname_1!!, phase(\loq)}{\loq'}$ and $\trans{\loq} {\actname_2!!, phase(\loq)}{\loq'}$ in $\php$. While \pairwise and internal transitions do not coordinate with other processes, transitions in \gspp are assumed to synchronize with all processes. As such, we create guards on these transitions based on phases, as we can be sure that if a process is in a phase to take such a transition, no process will be outside the phase. No feasible behaviors are removed.
Based on the simple translation from internal and \pairwise messages given above, the following correspondence between the global transitions based on those primitive holds:
\begin{equation}
\forall q_1,q_2 \in \GQ_{\merc}: \trans{q_1}{\actname}{q_2} \in \ml \implies
\trans{\absfn(q_1)}{\actname}{\absfn(q_2)} \in \mhp.  
\label{implication1}
\end{equation}

where, $\actname$ is a rendezvous or an internal action. Note that, for rendezvous transitions, we still need to show that the other direction holds. Since the \gspp does not support process indices, a global transition based on rendezvous communications in \aml programs between, say, process $p_1$ in state $\loq_1$ and process $p_2$ in state $\loq_2$ would correspond to a 2-sender transition involving \emph{any} process in $\loq_1$ and \emph{any} process in $\loq_2$. However, because $\pl$ is \chwellbehaved, the \pid-based communications are equivalent to the communication actions without \pids in the \gspp.


\para{\Pairwise Transitions under Full Symmetry.}
Recall that \aml syntax imposes syntactic constraints on how expressions of type $\Ind$ can be used. Essentially, 
the only expressions allowed are equality checks. These constraints ensure that the send statement \terma{\dslsend}(\actname, \nont{\optintvar},\nont{idExp}) (corresponding to the local send transition $\sendrzid{(\q,\sigma)}{\actname}{\pid}{(\q',\sigma')}$) cannot be instantiated with a specific, \emph{concrete} \pid. This \pid-independent behavior ensures that, if there exists a transition $\sendrzid{(\q,\sigma)}{\actname}{i}{(\q',\sigma')}$ for some $i \in \Ind$, then there exists a transition $\sendrzid{(\q,\sigma[i \gets j])}{\actname}{j}{(\q',\sigma'[i \gets j])}$ for all $j \neq i \in \Ind$. Hence, if a global transition $\trans{\gq_1}{\actname}{\gq_2}$ based on a \pairwise action $\actname$ between processes $P_i$ and $P_j$ exists in $\ml$, then $\forall \pi \in \PG. \ \trans{\pi(\gq_1)}{\actname}{\pi(\gq_2)}$ between processes $P_{\pi(i)}$ and $P_{\pi(j)}$ also exists in $\ml$.
%
%
%
%

\begin{lemma}
\label{lem:rendezvousfullsymppr}


Let $\actname$ be an rendezvous action 
and $\pl$ be a \chwellbehaved process. Then, $\forall q_1,q_2 \in \GQ_{\merc}: \trans{\absfn(q_1)}{\actname}{\absfn(q_2)} \in \mhp \implies \trans{q_1}{\actname}{q_2} \in \ml $.
\end{lemma}
\begin{proof}

Since permuting a global state of \aml programs does not change the local state of a process but just its \pid, we know that:  $\forall q_1,q_2 \in \GQ_{\merc}$ if $\ q_1 = \pi(q_2)$ for some $\pi \in \PG$ then $\absfn(q_1) = \absfn(q_2)$ (since the abstraction function $\absfn$ only captures the local state space, but not the indices). Since $\pl$ is rewritable, we know that there exists at most one \pairwise-receive transition per action per phase. Under full symmetry, this condition ensures that abstracting the receiver \pid (in a \pairwise-send transition $\sendrzid{s}{\actname}{\pid}{s'}$) does not introduce spurious behaviors. It then follows that:
$\forall q_1,q_2 \in \GQ_{\merc}: \trans{\absfn(q_1)}{\actname}{\absfn(q_2)} \in \mhp \implies \trans{q_1}{\actname}{q_2} \in \ml$.
\end{proof}
By~\lemref{rendezvousfullsymppr} and property~(\ref{implication1}) above, we obtain the following result:
\begin{lemma}
\label{lem:simulationEquiv}
Let $\actname$ be an rendezvous action 
and $\pl$ be a \chwellbehaved process. Then, $\forall q_1,q_2 \in \GQ_{\merc}: \trans{\absfn(q_1)}{\actname}{\absfn(q_2)} \in \mhp \iff \trans{q_1}{\actname}{q_2} \in \ml $.
%
\end{lemma}

\para{\textsc{\RewriteValueConsTransitions}.} We now show that $k$-sender actions in \gspp can be used to model \ValueCons \consagree. Consider a \ValueCons instance $\consnamev$ in $\pl$. Let $\chvar$ have the domain $V$. Let $W = \{ w_1,\ldots,w_m\}$ be the set of all sets constructed from elements in $V$ with size between 1 and $\chcard$ (inclusive). Also, for each $s_i \in src_\consnamev$, let $f_{i, j} \in \LQ$ denote the local state into which each participant in $s_i$ transitions after $l$ ends, for a given set of winning values $w_j \in W$. For each such $w_j$, create a $k$-sender  $a_j$ action as follows:

\begin{compactenum}[--]
\item create a transition $\trans{s_i}{a_j??,src_\consnamev}{f_{i, j}}$ in $\php$, and,
\item create for each element $e \in w_j$ a new local-send action $a_{j, e}!!$, and for each $s_i$ such that $\valOf{s_i}{v} = e$, create a transition $\trans{s_i}{a_{j, e!!},src_\consnamev}{f_{i, j}}$ to $\php$.
\end{compactenum}

\para{\textsc{RewritePartitionTransitions}} We now show how $k$-maximal actions in \gspp are used to model \PartitionCons \consagree. Consider a \PartitionCons instance $\consnamep$ in $\pl$. For the win transition $\trans{\ls}{\pcw(\chset,\chcard)}{\ls’}$ in $\pl$, we create the transitions $\trans{\ls}{a_i!!,src_\consnamep}{\ls’} \ \forall i \in \{1\ldots k\}$ in $\php$, where $a$ is a $k$-maximal action. For each $\trans{\ls}{\pcw(\chset,\chcard)}{\ls’}$ in $\pl$ we create the receive transition $\trans{\loq}{a??, src_\consnamep}{\loq'}$ in $\php$.

We show that the \textsc{RewritePartitionTransitions} and \textsc{\RewriteValueConsTransitions} procedures described above yield an equivalence between agreement transitions and the corresponding $k$-sender and $k$-maximal transitions used to model them.

\begin{lemma}
\label{lem:mappingChooseTrans1}
Let $\trans{\gq_1}{}{\gq_2}$ be an \PartitionCons agreement transition with cardinality $k$, $b$ be a $k$-maximal action, and $G$ be a global guard built as described by the \textsc{RewritePartitionTransitions} in~\algoref{mapping}. Then,
$\forall \gq_1, \gq_2 \in \GQ_{\merc}: \trans{\gq_1}{}{\gq_2} \in \ml \iff \trans{\absfn(\gq_1)}{b, G}{\absfn(\gq_2)} \in \mhp$
\end{lemma}
\begin{proof}
The semantics of the agreement transition in \aml is to pick up to a total of $k$ winning participating processes from multiple possible start states $\{c_1,\ldots,c_m\}$ and switch their states to the corresponding winning states $\{w_1,\ldots,w_m\}$ while all the other participating processes move to the corresponding losing states $\{l_1,\ldots,l_m\}$. 

A $k$-maximal transition in \gspp behaves similar to a agreement transition by moving $k$ processes from their respective $c_i$ states to a winning state $w_i$ and all the other processes to a losing state $l_i$. Since we create a $k$-maximal transition with the same number of winning' send transitions for each starting start $c_i$, winning processes may be arbitrarily distributed among the starting states, similar to the corresponding \aml transition. In both models, the rest of the processes move to the losing states. 

Hence, the semantics of $k$-maximal transitions enforces the \emph{Consistent Winners} post condition of the corresponding agreement transition. The guard $G$ ensures that the broadcast is only enabled when all the processes participating in the \consagree round are in the right states, hence, ensuring the \emph{Consistent Participants} precondition. Since the guard $G$ is a set of local states, it is invariant to the abstraction function $\absfn$ (i.e. $G$ is enabled in $\gq$ iff it is enabled in $\absfn(\gq)$).
\end{proof}

\begin{lemma}
\label{lem:mappingChooseTrans2} Let $\trans{\gq_1}{}{\gq_2}$ be a \ValueCons transition with cardinality $k$, $b$ be a $k$-sender transition, and $G$ be a global guard built as described by the \textsc{\RewriteValueConsTransitions} in~\algoref{mapping}. Then,
$\forall \gq_1, \gq_2 \in \GQ_{\merc}: \trans{\gq_1}{}{\gq_2} \in \ml $
$\iff \trans{\absfn(\gq_1)}{b, G}{\absfn(\gq_2)} \in \mhp$
\end{lemma}
\begin{proof}
The semantics of the agreement transition in \aml is to pick up to $k$ winning values proposed by the participating processes starting from multiple possible start states $\{c_1,\ldots,c_m\}$ and switch the processes' states to the corresponding next states where \consagree is reached on a set of $1$ to $k$ values.

By construction, each \ValueCons transition is simulated by at least one $k$-sender transition that guarantees agreement on a given set of winning values. The semantics of the generated $k$-sender transition ensures that it is \emph{only} enabled when there exists at least one participant proposing each value of the set of winning values. This is achieved by placing the local-send transitions from each state $s$ with $\valOf{s}{v} \in w$, where $w$ is the winning set. As in~\lemref{mappingChooseTrans1}, the guard $G$ ensures that the transition is only enabled in a state satisfying \emph{Consistent Participants} precondition.
\end{proof}
\para{\textsc{RewriteCrashTransitions}.} Since process crashes are nondeterministic, they behave like internal transitions, and are rewritten similarly.

\para{Simulation Equivalence of $\ml$ and $\mhp$.} Based on  Lemmas \ref{lem:broadcasts} - \ref{lem:mappingChooseTrans2}, we obtain the following theorem:

\begin{theorem} \label{thm:equiv} 
Given a rewritable \chwellbehaved $\pl$, \\$\php =$ \Call{Rewrite}{$\pl$}, 
the respective global transition systems 
 $\ml(n)$, $\mhp(n)$, and \permissible 
specification $\phi$, we have:
\vspace{-1em}

 $$\forall n. \; \mhp(n) \models \phi  \iff  \ml(n) \models \phi.$$
\vspace{-1.5em}

\end{theorem}


%
%

%


We now show that, for any rewritable \chwellbehaved \machine $\pl$, the parameterized verification of $\pl$ for a \permissible specification $\spec{m}{f}{n}$ is decidable. We show that $\php$ = \Call{Rewrite}{$\pl$} falls within the decidable fragment of the \gspp.

\para{Sufficient Well-Behavedness Condition.} First, we define a condition over \gspp processes which generalizes the well-behavedness conditions for \gspp. For a $k$-sender or $k$-maximal action $a$, let $\senderset$ be the set of states $\ls_i$ such that $\trans{\ls_i}{a_i!!, \guard}{\ls_i'}$, $1 \leq i \leq k$, are the local sending transitions, $\senderset'$ the set of all states $s_i'$, and $M_a(\ls_k)$ denote the state $\ls_l$ that $\ls_k$ is mapped 
to in a transition $\trans{\ls_k}{a??}{\ls_l}$. 
We say that action $a$ is 
\emph{weakly guard-compatible} if the following holds for all $\guard' \in \guardset$, the set of all guards in the system:
%
%
%
\begin{multline}
\forall (a1,G1),(a2, G2)\in \mathcal{AG}. 
(\senderset_{a1}' \cap G2 = \varnothing) \lor \\ ((\senderset_{a1}' \cup M_{a1}(G1))\cap \senderset_{a2} = \varnothing) \lor (\senderset_{a1}' \subseteq G2 \land  \forall s \in G1. \exists t \in G2. M_a(s) \rightsquigarrow t)
\tag{wellBehaved} \label{eq:big hammer}
\end{multline}

where $M_a(\ls) \rightsquigarrow t$ represents that the receiver must have an unguarded path of internal transitions to the state $t$ from $M_a(\ls)$. Informally, this condition requires that all senders of a transition must end outside of a guard together, all process performing $a1$ end outside of the send set of $a2$, or all processes must end inside the guard. We point out that this condition implies all of the well-behavedness conditions of \gspp.

\begin{claim}
\\$\forall\ \pl.\ \text{\chwellbehaved}(\pl) \rightarrow \text{wellBehaved}(\php)$, where $\php = \Call{Rewrite}{\pl}$.
\end{claim}

\begin{claimproof}
This will be a proof by contradiction. Assumption: $\exists\ \pl.\  \text{\chwellbehaved}(\pl) \land \neg \text{wellBehaved}(\php).$ More specifically, the following must be true for $\php= \Call{Rewrite}{\pl}$. 
%
\begin{multline}
	\exists (a1,G1),(a2, G2)\in \mathcal{AG}. 
	(\senderset_{a1}' \cap G2 \not = \varnothing) \land \\
	((\senderset_{a1}' \cup M_{a1}(G1))\cap \senderset_{a2} \not = \varnothing) \land
		(\senderset_{a1}' \not \subseteq G2 \lor 
		\exists s \in G1. \neg(\exists t \in G2. M_a(s) \rightsquigarrow t))
\end{multline}
%

That is, for some guard $G2$ and action $a1$,
some acting transition on $a1$ ends in $G2$, some process may perform $a1$ and end in a state with an acting transition on $a2$, and either an acting transition on $a1$ ends outside of $G2$, or a receiver cannot reach $G2$.


\paragraph{Globally-Synchronizing Actions}
Consider the case in which $a1$ is a globally-synchronizing action. 


In the scenario where some acting transition ends outside $G2$, we examine the two possibilities under which the guard $G2$ may have been created.
\begin{compactenum}
\item If $G2$ is $src_{a2}$ for some $a2 \in \gsactions$, we have a contradiction, because $\text{\chwellbehaved}(\pl)$ ensures that 
if an acting transition of a globally-synchronizing action (e.g., $a1$) ends in a state with an outgoing reacting action (e.g., $a2$), and $a2$ is \firable in $dst_{a1}$, then all other acting transitions of $\act$ must end in a state with an outgoing receive action $a2$, which is in $src_{a2}$.
\item If $G2$ is the guard of an internal or pairwise transition, $G2$ is a union of phases. Since any two states in any destination set of a globally-synchronizing action always appear together in phases, we have a contradiction with the assumption that one of them ends outside of $G2$.
\end{compactenum}
We proved that acting transitions must end in $G2$ if any of them does. Now, consider the scenario where the acting transitions end inside $G2$ but some reacting transition cannot reach $G2$ (via some path of unguarded internal transitions). We again examine the two creation scenarios of $G2$.
\begin{compactenum}
\item If $G2$ is $src_{a2}$ for some $a2 \in \gsactions$, we have a contradiction since our
restriction on reacting actions imposed by $\text{\chwellbehaved}(\pl)$ ensures that because $a2$ is \firable in $dst_{a1}$, then the reacting transition's destination state must be able to reach a source state of $a2$. 

\item If $G2$ is the guard of an internal or pairwise transition, $G2$ is a union of phases. Since any two states in any destination set of a globally-synchronizing action always appear together in phases, we have a contradiction with the assumption that one receiver ends outside of $G2$.

%

\end{compactenum}
Thus, globally-synchronizing actions in $\pl$ map to guard-compatible actions in $\php$.

\paragraph{Internal}
Consider the case in which $a1$ is an internal action. This is a transition in $\php$ with only one sending transition, so there must be some self-looping receive transition which cannot reach $G2$. This requires that, for $G1$ (the guard on the internal transition), $G1 \not = G2$. We consider the two possibilities under which the guard $G2$ may have been created.
\begin{compactenum}
\item If $G2$ is $src_{a2}$ for some $a2 \in \gsactions$, then the state of the self-looping receive transition (which is in $G1$) must not have a path of internal transitions to reach a source state of $a2$. However, our $\text{\chwellbehaved}(\pl)$ conditions on internal actions state that, if $a2$ is \firable in $G1$, then every state in $G1$ has a path to a state with a receive of $a2$, so we have a contraction.

\item If $G2$ is the guard of an internal or pairwise transition (i.e. some $\phase$), then the state of the self-looping receive transition (a state in $G1$, which is also a phase) must not have a path of internal transitions to reach $G2$. However, the sender begins in $G1$ and ends in $G2$, so by our definition of phases, $G1$ and $G2$ would be merged, so $G1$ = $G2$, and no such self-looping transition can exist. So, we have a contradiction.
\end{compactenum}
Thus, internal actions in $\pl$ map to guard-compatible actions in $\php$.
\paragraph{\Pairwise}
Consider the case in which $a1$ is a \pairwise action. There are two senders, so one sender must end in some guard $G2$ and either the other sender must end outside of $G2$, or some receiver must not be able to reach $G2$.
Considering the scenario where some sender (i.e. the \pairwise sender or receiver) ends outside $G2$, we consider the two creation scenarios of $G2$.
\begin{compactenum}
\item If $G2$ is $src_{a2}$ for some $a2 \in \gsactions$, we have a contradiction, because $\text{\chwellbehaved}(\pl)$ ensures that if $a2$ is \firable in $G1$ and either the \pairwise sender or receiver ends in a state where there is a globally-synchronizing reacting transition on $a2$, then the other must also be able to reach it, which contradicts our assumption. 
\item If $G2$ is the guard of an internal or pairwise transition, we also have a contradiction, because all source and destination states of a \pairwise action will be in the same phase.
\end{compactenum}
Now that we know the senders of the 2-sender action must both end in $G2$ if either of them does, consider the scenario where some self-looping receiver ends outside $G2$. We consider the two creation scenarios of $G2$.
\begin{compactenum}
\item If $G2$ is $src_{a2}$ for some $a2 \in \gsactions$, then the state of the self-looping receive transition (which is in $G1$) must not have a path of internal transitions to reach a source state of $a2$.
However, $\text{\chwellbehaved}(\pl)$ ensures that, because $a2$ is \firable in $G1$, this is not the case, so we have a contradiction. 
\item If $G2$ is the guard of an internal or pairwise transition (i.e. some $\phase$), then the assumption requires that  the state of the self-looping receive transition (which is in $G1$) must not have a path of internal transitions to reach $G1$.  However, both senders begin in $G1$ and ends in $G2$, so by our definition of phases, $G1$ and $G2$ would be merged, so $G1$ = $G2$, and no such self-looping transition can exist. So, we have a contradiction.
\end{compactenum}
Thus, \pairwise actions in $\pl$ map to guard-compatible actions in $\php$.

Since all transitions in the transitions in $\pl$ map to guard-compatible transitions in $\php$, $\php$ is well-behaved, contradicting the assumption.
\end{claimproof}

\section{Cutoffs for Efficient Parameterized Verification}
\label{app:cutoffapp}
We show that, for any \amenable \machine $\pl$ and \permissible specification $\spec{f}{m}{n}$ has a cutoff $c =  m$.
We show that when any such \amenable \machine $\pl$ is mapped to $\ph = \Call{Rewrite}{\pl}$ that is guaranteed to have the cutoff $c = m$. 
The \amenability conditions (ref. \secref{cutoffs}) assume \permissible specifications of the form $\spec{m}{f}{n}$. For simplicity, we assume the existence of a state $s$ such that each local state in $\states(f)$ has an internal transition to $s$. We write $\cutoffch(\pl)$ to denote $\pl$ satisfies the cutoff amenability conditions and $\cutoffgsp(\php)$ to denote that the cutoff conditions in \gspp hold for process $\php$. A transition of a \gspp process $\php$ is \emph{\free} if it is (i) an internal transition, (ii) a sending transition of either a broadcast (i.e., a 1-sender action) or a $k$-maximal action, or (iii) a receiving transition $\trans{\ls}{a??,\guard}{\ls'}$ of a broadcast with matching sending transition $\trans{\ls}{a!!,\guard}{\ls'}$. A path from one state to another is \free if all transitions on the path are \free.

We now show that if a process $\pl$ satisfies the \amenability conditions, then the corresponding $\php$ satisfies $\cutoffgsp$, and thus $\pl$ has cutoff $c = m$.\ourskip

\begin{claim} 
$\cutoffch(\pl) \rightarrow \cutoffgsp(\php)$, where $\php = \Call{Rewrite}{\pl}$.
\end{claim}

\begin{claimproof}
This will be a proof by contradiction, assume: $\cutoffch(\pl) \land \neg\cutoffgsp(\php)$.

In order for $\cutoffgsp(\php)$ to not hold, 

\begin{compactenum}
\item there must exist a path from $\ls_0$ to $\ls$ which is not \free, and,
\item there must exist a transition $\trans{s_s}{}{s_d}$ starting on a free path ($s_s \in \fgsp$) and leaves it ($s_d \not\in \fgsp$)
  and either

\begin{compactenum}
\item there are no paths out of $s_d$,
\item some path from $s_d$ that does not lead back to $\ls_s$, or
\item some path from $s_d$ that leads back to $\ls_s$ and is not free between $s_d$ and $\ls_s$.
\end{compactenum}

\end{compactenum}
In case (1) where
all paths from $\ls_0$ to $\ls$ are \freech, there must be some \freech transition in $\pl$ which maps to a transition which is not \free.
This  transition must be an acting \ValueCons transition, because winning \PartitionCons, broadcast send, and internal transitions all map directly to \free transitions.
In the case where the cardinality $\chcard$ is 1, such a transition maps to a broadcast send transition $t$, as well as a corresponding receive $u$, which are \free transitions because they form a negotiation.
The transition also maps to other broadcast receive transitions corresponding to sends from other states where the decided value is the same as that of $t$.
Since taking any of these receive transitions is
semantically indistinguishable from taking the $u$ transition, we
consider
these additional receives
to be
part of the negotiation and thus are equally \free. As such, all \freech transitions map to \free transitions, so all paths in $\fgsp$ are \free if all paths in $\fch$ are \freech.
Thus, we arrive at a contradiction, because all paths from $\ls_0$ to $\ls$ are \freech, so there cannot be a path from $\ls_0$ to $\ls$ which is not \free.

%
%
%
%
%
%


If the second \amenability condition holds, 
we get a contradiction as follows.

Some transition $\trans{s_s}{}{s_d}$ that starts on a free path ($s_s \in \fgsp$) and leaves it ($s_d \not \in \fgsp$), must be the result of rewriting some transition in $\pl$.
According to our \amenability conditions, all paths out of $s_d$ must lead back to $s_s$ and be \freech between $s_d$ and $s_s$, and such a path must exist. Since every state in $\fch$ is a state in $\fgsp$ (ref. \lemref{lamma9}), then $s_s$ is in $\fgsp$ and the path from $s_d$ to $s_s$ is free, so we have a contradiction.

%
%
%
%
%
%

Since we arrive at a contradiction on all proof branches, if our \amenability conditions hold on $\pl$, then the cutoff conditions of \gspp hold on $\php$.

\end{claimproof}



While the definitions of $\fch$ and $\fgsp$ differ, it can be shown that the states they include are the same. Note that the local state spaces of $\pl$ and $\php$ are the same. Let $states(\mathcal{F}) = \{ s \in p \mid p \in \mathcal{F}\}$ be the set of states of all the paths in $\mathcal{F}$.

\begin{lemma}
For any $\pl$ such that $\php$ = \Call{Rewrite}{$\pl$}, $states(\fch) = states(\fgsp)$. 
\label{lem:lamma9}
\end{lemma}
\begin{proof}

For any path $p \in \fch$, if there are no \ValueCons transitions in $p$, then all transitions in $p$ are \freech and map to free transitions in $\php$. Hence, all states in $p$ will be in $\fgsp$. 
Consider a scenario where two sets of free paths $b$ and $a$ are connected by a \valstore transition $t_v$. Each path $b_i \in b$ begins at a state $first(b_i)$ and ends at a state $last(b_i)$, which is a source state of $t_v$. Each path $a_j \in a$ begins at a destination state of $t_v$, $first(a_j)$, and ends at a state $last(a_j)$. 

After rewriting, every source state $last(b_i)$ of $t_v$ has a 1-sender broadcast-send transition from $last(b_i)$ to some $first(a_j)$, and every destination state $first(a_j)$ of $t_v$ is reached by 1-sender broadcast-send transition from some $last(b_i)$. Therefore, every $first(b_i)$ has a free path to some $last(a_j)$ through the 1-sender broadcast-send from $last(b_i)$ to $first(a_j)$, and every source and destination state of $t_v$ is on at least one of these free paths. 

The paths in $\fch$ can be viewed as a series of sets of free paths separated by \valstore transitions. So, by applying the above logic repeatedly, we can see that every source and destination state of every \valstore transition in $\fch$ is, in fact, part of a free path in $\fgsp$.
\end{proof}

\subsection{Reachability of Combinations of States.}
\label{app:generalSpecs}
%
\begin{newtext}
We handle conjunction and disjunction of the \permissible specification $\spec{m}{f}{n}$ as follows.
Consider the correctness property of the form:
$$\phi_{m_1,f_1}(n)  \wedge \ldots \wedge \phi_{m_j,f_j}(n) $$
for some finite $j$. 
Then, we compute the cutoff $m_j$ independently for each $j$, and use the maximum cutoff value $m = max(m_1,\ldots,m_j)$ to check the correctness of the system. A system that satisfies this property is one that satisfies all conjuncts. Subsequently, a global error state for this property is one that violates \emph{any} conjunct, so one can check each conjunct individually with the corresponding $m_j$. Hence, it is sufficient to check the correctness of the system with the maximum cutoff since that system has a sufficient number of processes to produce any reachable error states in a system with fewer than $m$ processes.

Next, consider the correctness property of the form:
$$\phi_{m_1,f_1}(n)  \lor \ldots \lor \phi_{m_j,f_j}(n) $$
for some finite $j$.
Then, we compute the cutoff $m_j$ independently for each $j$, and use the sum of the resulting cutoff values $m = sum(m_1,\ldots,m_j)$ to check the correctness of the system. 
A system that satisfies this property is one that always satisfies some disjunct. 
Subsequently, a global error state for this property is one that violates \emph{all} disjuncts, so one must check correctness of a system where all the processes needed to reach a global error state are available in the respective local error states (characterized by $f_1,\ldots,f_j$) at the same time. For each $i \in [1,j]$, let $\mathcal{F}_i$ denote the set of paths from the initial state $\ls_0$ to $\states(f_i)$. To ensure that one can compute the cutoffs independently, we require that for any two sets of paths $\mathcal{F}_{i_1}$ and $\mathcal{F}_{i_2}$ there does not exist any non-internal transition $\trans{s}{}{s'}$ such that $s \in states(\mathcal{F}_{i_1})$ and $s' \in states(\mathcal{F}_{i_2})$.

Finally, cutoff values to check correctness properties that are conjunction or disjunction (but not negation) of permissible safety specifications can be computed as follows. First, one can convert the property to CNF form. Then, apply the disjunct rule to the clauses and the conjunct rule to the resulting simplified formula.
\end{newtext}

\end{document}